\documentclass[aps,prb,twocolumn,showpacs,superscriptaddress,groupedaddress]{revtex4}
\usepackage{amsmath,amssymb,color,graphicx,hyperref,amsthm,bm,epstopdf,mathrsfs}
\usepackage{hyperref}
\usepackage{graphicx,grffile}
\usepackage{amsmath}
\usepackage{amsfonts}
\usepackage{array}
\usepackage{url}
\usepackage{flexisym}
\usepackage{comment}
\usepackage[normalem]{ulem}

\usepackage{subcaption}
\usepackage{float}

\usepackage[export]{adjustbox}
\usepackage[toc,page]{appendix}

\renewcommand{\ol}{\overline}

\renewcommand{\eqref}[1]{Eq. \ref{#1}}
\newcommand{\figref}[1]{Fig. \ref{#1}}

\newcommand{\secref}[1]{Sec. \ref{#1}}

\renewcommand{\AA}{\mathcal{A}} 
\newcommand{\BB}{\mathcal{B}} 
\newcommand{\CC}{\mathcal{C}} 
\newcommand{\Q}{\mathbb{Q}} 
\newcommand{\R}{\mathbb{R}} 
\newcommand{\eps}{\varepsilon}
\newcommand{\reK}{\R}
\newcommand{\imK}{\mathbb I}

\newcommand{\nvec}{\mbox{\boldmath{$n$}}}
\newcommand{\im}{\mbox{Im}}

\newcommand{\nv}{\nvec}

\definecolor{TW-color}{RGB}{100,0,100}
\definecolor{AW-color}{RGB}{0,0,100}
\definecolor{Error-color}{RGB}{250,0,0}

\graphicspath{{./images/}}

\makeatletter
\let\cat@comma@active\@empty
\makeatother

\begin{document}

\title{Magnetic field behaviour in $s+is$ and $s+id$ superconductors: twisting of applied and spontaneous fields.}

\author{Martin Speight}
\affiliation{School of Mathematics, University of Leeds, Leeds LS2 9JT, United Kingdom}
\author{Thomas Winyard}
\affiliation{School of Mathematics, University of Leeds, Leeds LS2 9JT, United Kingdom}
\author{Alex Wormald}
\affiliation{School of Mathematics, University of Leeds, Leeds LS2 9JT, United Kingdom}
\author{Egor Babaev}
\affiliation{Department of Physics, KTH-Royal Institute of Technology, Stockholm, SE-10691 Sweden}

\begin{abstract}
We consider magnetic field screening and spontaneous magnetic fields in $s+is$ and $s+id$ superconductors both analytically and numerically.
We show that in general, the linearized model couples the moduli of order parameters to the magnetic modes. This causes magnetic field screening that does not follow the standard exponential law and hence cannot be characterized by a single length scale: the London penetration length.

We also demonstrate that the resulting linear mixed modes, correctly predict  spontaneous fields and their orientation. We show that these mixed modes cause external fields to decay non-monotonically in the bulk. This is observed as the magnetic field twisting direction, up to an angle of $\pi/2$, as it decays in the nonlinear model. 

Finally, we demonstrate that there are two non-degenerate domain wall solutions for any given parameter set. These are distinguished by either clockwise or anti-clockwise interpolation of the inter-component phase difference, each producing a different solution for the other fields. However, only domain wall solutions in $s+id$ systems exhibit magnetic field twisting.
\end{abstract}

\maketitle

\section{Introduction}

Recent experiments have reported the discovery
of an $s+is$ superconducting state
in  Ba$_{1-x}$K$_x$Fe$_2$As$_2$ \cite{grinenko2020superconductivity,grinenko2017superconductivity,grinenko2021bosonic}. Such spin-singlet pairing states, that spontaneously break time reversal symmetry, have long been predicted
  \cite{ stanev2010three,carlstrom2011length,maiti2013s+,boker2017s+,ChubukovLiFeAs,Hirschfeld2015,kreisel2020remarkable}, along with the related  $s+id$  states
\cite{lee2009pairing,Khodas2012,platt2012mechanism},  to form in multi-band superconductors. In an effective model, these are described by at least two complex fields or order parameters.

Such states can host multiple interesting phenomena, such as massless modes \cite{carlstrom2011length,Lin2012}, mixed collective modes \cite{carlstrom2011length,Stanev2012,marciani2013legett,Maiti2013, silaev2018non, garaud2018properties,xu2020superconducting,maiti2015collective,muller2018short}, new flux flow phenomena \cite{Silaev2013a}, stable and metastable
Skyrmions \cite{Garaud2011,Garaud2013,winyard2018skyrmion}, new thermoelectric effects \cite{Silaev.Garaud.ea:15,Garaud2016} and new fluctuation-induced phases
\cite{bojesen2013time,bojesen2014phase,carlstrom2014spontaneous,grinenko2021bosonic}.

Both $s+is$ and $s+id$ systems are characterized not only by spontaneous breakdown of time reversal symmetry (BTRS) but also by the appearance of non-collinear gradients of the inter-component phase difference and relative densities around impurities \cite{garaud2014domain,vadimov2018polarization,grinenko2020superconductivity,benfenati2020magnetic}.

The nature of spontaneous magnetic fields near impurities in these systems is different from that in chiral systems, such as in $p+ip$ superconductors \cite{sigrist1991phenomenological,bouhon2014current,speight2019chiral}. 
Spontaneous magnetic fields in $s+is$ superconductors are more subtle, and their existence has been a subject of recent debate \cite{ovchinnikov2019singular,silaev2019comment}.
It has also recently been suggested that spontaneous fields around impurities exist for $s+id$ systems \cite{lee2009pairing,lin2016distinguishing,vadimov2018polarization,Garaud2016}, due to the non-collinear gradient terms.

In contrast to the better studied $p+ip$ systems,
the  spontaneous fields generated in $s+is$
superconductors have only recently started to be explored \cite{ChubukovMaitiSigrist,lin2016distinguishing,Silaev.Garaud.ea:15,Garaud2016,garaud2018properties,benfenati2021spontaneous}. In particular in \cite{benfenati2020magnetic} a comparative study was presented of the magnetic fields generated by domain walls in both $s+is$ and $s+id$ superconductors.

 Note that, we will also use ``spontaneous magnetic field'' to refer to fields generated in response to applied external field ${\bf H}$ but in a direction perpendicular to ${\bf H}$.

To demonstrate why anisotropic BTRS $s$-wave systems have such different properties, consider an ordinary superconductor, with a single order parameter $\psi$ and no crystal anisotropies. The system is well described by the London model, exhibiting exponential decay of both the magnetic field $\boldsymbol{B}$ and the matter field $|\psi|$ (order parameter magnitude) away from a defect in the superconducting state. This exponential decay is governed by the London penetration depth $\lambda$ and coherence length $\xi$ respectively, \cite{landau1950k,Tinkham1995,Svistunov2015}
\begin{equation}
\boldsymbol{B} = \boldsymbol{B}_0 e^{-r/\lambda}, \quad |\psi| = u - |\psi_0|e^{-r/\xi},
\end{equation}
restoring the fields to their ground state value $(\boldsymbol{B},\psi) = (\boldsymbol{0},u)$. 

Introducing an additional order parameter to an ordinary superconductor creates a two-component isotropic system. This system also exhibits exponentially decaying physical quantities. This decay is govern by a London penetration depth $\lambda$ and two coherence lengths $\xi_\alpha$, one for the magnitude of each component $|\psi_\alpha|$, as well as an additional Leggett mode for the phase difference between the two complex order parameters.

Most superconducting materials are anisotropic.
Multiband $s+is$ and $s+id$  systems  exhibit anisotropy in each band, which can be calculated from the symmetries of the associated Fermi surface. If there are  non-trivial inter-component gradient couplings in a time-reversal-invariant system, then the London and Leggett modes in general hybridize \cite{silaev2018non}. This leads to the magnetic field and phase difference coupling, such that each of the quantities decays as two competing exponentials with different length scales. This can lead to non-trivial vortex states or Skyrmions \cite{winyard2018skyrmion,winyard2018}. However, if
 time reversal symmetry is broken then all modes are generically coupled, including the order parameter magnitudes. For example, in a $p+ip$ superconductor in an inhomogeneous state, solutions for each physical field in general are described by all of the anisotropic length scales \cite{speight2019chiral}.
This complexity motivates the systematic investigation of magnetic properties of $s+is$ and $s+id$ superconductors.

In this paper we will study an effective Ginzburg-Landau (GL) model for  $s+is$ and $s+id$ pairing symmetries.  We will expand previous studies of anisotropy effects, demonstrating that such systems can only be described by anisotropic mixed modes. Using this we will make two key experimentally verifiable predictions for $s+is$ and $s+id$ systems:
\begin{itemize}
\item Magnetic field twisting - the mixed modes predict that the magnetic field will twist direction when decaying from a defect. 
\item Spontaneous magnetic field - fluctuations in the matter fields, due to coupled linear modes, must excite fluctuations in the magnetic field.
\end{itemize}

Hence, excitations that are commonly associated with purely the matter fields, such as domain walls and defects,  in $s+is$ and $s+id$ models will exhibit a spontaneous magnetic response. This confirms previous numerical calculations that have been performed for domain walls \cite{benfenati2020magnetic}. It has been suggested that defects do not produce spontaneous magnetic field in such models \cite{ovchinnikov2019singular}. However, the work in this paper supports the authors previous comment on this suggestion \cite{silaev2019comment}. 

We will perform numerical simulations of both the Meissner state and domain walls, comparing the results with the predictions of the linear modes. In particular, we will demonstrate magnetic field twisting and spontaneous fields for both. 
 
\section{Anisotropic 2-Component  Model}
We consider a multiband dimensionless anisotropic Ginzburg-Landau (GL) free energy,
\begin{align}
\label{Eq:F}
F = \int_{\mathbb{R}^3}  \left( \frac{1}{2}Q_{ij}^{\alpha \beta} (D_i \psi_\alpha)^*D_j\psi_\beta +\frac{\left(\nabla \times A \right)^2}{2} + F_P \right),
\end{align}
where we have used Greek indices to denote components of the order parameter $\psi_\alpha$ and Latin indices for spatial directions. Repeated indices will denote summation throughout.
Such models can be microscopically derived (e.g. in \cite{garaud2017microscopically}). We are interested  in 2-component models, thus the order parameter for the condensate is represented as two complex fields,
\begin{equation}
\psi_\alpha = \rho_\alpha e^{i\theta_\alpha}
\end{equation}
where $\alpha \in \{1,2\}$. As GL theory is a $U(1)$ gauge theory, we include a gauge field $A_i$ and corresponding covariant derivative $D_i = \partial_i - i A_i$. The gauge invariant magnetic field is then  $B_k = \varepsilon_{ijk}\partial_iA_j$. We find the GL field equations by taking the variation of \eqref{Eq:F} with respect to the fields $\psi_\alpha$ and $A_i$,
\begin{align}
Q^{\alpha\beta}_{ij}D_i D_j \psi_\beta &= 2 \frac{\partial F_p}{\partial \overline{\psi}_\alpha}  \label{Eq:GLeom1} \\ 
-\partial_j(\partial_j A_i - \partial_i A_j ) &= \im(Q^{\alpha\beta}_{ij} \overline{\psi}_\alpha D_j \psi_\beta),
\label{Eq:GLeom}
\end{align}
where \eqref{Eq:GLeom} is the anisotropic version of Amp\`ere's Law and thus we define the right hand side of this equation to be the supercurrent $J_i$. 

The gradient term in \eqref{Eq:F} is positive definite, hence the ground state solutions are the constant configurations that globally minimise $F_p$. As the potential term must be gauge invariant, it can only depend on the condensate magnitudes $\rho_\alpha$ and the phase difference $\theta_{12} := \theta_1 - \theta_2$. The phase difference terms will determine the symmetry of the target space, where BTRS ground states exhibit spontaneous symmetry breaking to a $U(1)\times Z_2$ symmetry. We choose the simplest BTRS term,
\begin{equation}
F_p = V(\rho_1, \rho_2) + \frac{\eta}{8}\rho_1^2 \rho_2^2 \cos 2\theta_{12},
\label{eq:Fp}
\end{equation}
where $\eta > 0$. This choice for the potential leads to a degenerate ground state, corresponding to two gauge inequivalent solutions $\theta_{12} = \pm \pi/2$. The remaining potential terms are assumed to be of the traditional form,
\begin{equation} \label{eqn:V}
V(\rho_1, \rho_2) = \alpha_\alpha \rho_\alpha^2 + \frac{\beta_\alpha}{2} \rho_\alpha^4 + \gamma \rho_1^2 \rho_2^2,
\end{equation}
where $\alpha_\alpha < 0$, $\beta_\alpha > 0$ and $\eta/8 - \gamma < \beta_\alpha$ so that the non-zero minimum value of $\rho_\alpha = u_\alpha > 0$ and both condensates are superconducting. A direct consequence of the $Z_2$ degeneracy in the ground state is the existence of domain wall solutions. These 1-dimensional defects occur when the phase difference interpolates between the two disconnected ground state values, $\theta_{12} = \pm \pi/2$, forming a 2-dimensional wall in the order parameter. 

The difference between this system and a standard multi-component GL model is the anisotropy matrices $Q^{\alpha\beta}$. To ensure that the energy is real they must satisfy $Q^{\alpha\beta}_{ij} = \overline{(Q^{\beta\alpha}_{ji})}$.

Note that for an $s+is$ or $s+id$ system, the form of these matrices can be derived from a microscopic model, by starting with a clean 3 band model, relevant for iron based compounds. It has been shown that under certain conditions a three-band model is described
by the above two component GL model \cite{garaud2017microscopically}. The form of the anisotropy matrices is derived from the symmetries of the Fermi surface (see e.g. \cite{garaud2016microscopically,vadimov2018polarization}) and is given in table \ref{Tab:Q}.  

\begin{table}
\begin{center}
\begin{tabular}{c|c} 
 \hline
 s+is & s+id\\
 \hline
 $Q^{11} = \left(\begin{array}{ccc} a_1 & 0 & 0 \\ 0 & a_1 & 0 \\ 0 & 0 &b_1 \end{array}\right)$ &  $Q^{11} = \left(\begin{array}{ccc} a_1 & 0 & 0 \\ 0 & a_1 & 0 \\ 0 & 0 &b_1 \end{array}\right)$\\
 $Q^{22} = \left(\begin{array}{ccc} a_2 & 0 & 0 \\ 0 & a_2 & 0 \\ 0 & 0 &b_2 \end{array}\right)$ &  $Q^{22} = \left(\begin{array}{ccc} a_2 & 0 & 0 \\ 0 & a_2 & 0 \\ 0 & 0 &b_2 \end{array}\right)$\\
  $Q^{12} = \left(\begin{array}{ccc} a_3 & 0 & 0 \\ 0 & a_3 & 0 \\ 0 & 0 &b_3 \end{array}\right)$ &  $Q^{12} = \left(\begin{array}{ccc} a_3 & 0 & 0 \\ 0 & -a_3 & 0 \\ 0 & 0 &b_3 \end{array}\right)$\\
 \hline
\end{tabular}
\end{center}
\caption{General form of the anisotropy matrices for $s+is$ and $s+id$ systems, derived from a microscopic model of a clean 3 band iron based system. \cite{garaud2017microscopically}}
\label{Tab:Q}
\end{table}

The $s+is$ matrices exhibit a continuous $SO(2)$ symmetry about the $z$-axis. They also exhibit an additional $C_2$ 2-fold symmetry about the $x,y$ axes giving a symmetry of $C_2 \times C_2 \times SO(2)$. In contrast the $s+id$ model has only a 2-fold symmetry in the basal $(x,y)$ plane, leaving the system with just a $C_2\times C_2 \times C_2$ symmetry.

\section{Linearized Model}\label{sec:lin}
We now consider the spatial dependence of fields decaying far from some defect.  Generally this is governed by the nonlinear GL equations \eqref{Eq:GLeom}, which must be solved numerically. However, fields are observed to decay to their ground state values far from a given excitation. Hence, we can approximate the long range behaviour of excitations by assuming that the fluctuations of fields about their ground state values is small, linearizing the equations of motion.

The standard approach is to consider each field individually, expanding the field about its ground state value while keeping all others constant. In the standard GL model, this leads to the famous London model for fluctuations in the magnetic field $\boldsymbol{B}$ and a separate matter equation for perturbations in the single condensate magnitude $|\psi|$. Whether the superconductor is of type I or type II can then be determined by which of these has the longer length scale. However, the correct derivation of  this result should be to linearize all fields together, showing that in the linear limit the magnetic and matter equations of motion decouple.

These two approaches ultimately lead to the same result for a single component superconductor. However, for a multicomponent anisotropic model it has been shown that the magnetic and matter equations do not in general decouple in the linear limit \cite{silaev2018non,speight2019chiral}. Hence, we cannot rely on the London model to describe the magnetic response of our system and must expand around all quantities simultaneously. 

We will first write our energy functional in terms of gauge invariant quantities. To achieve this we introduce a new gauge invariant vector field,  
\begin{equation}
p_i := A_i - \partial_i\theta_\Sigma,\qquad \theta_\Sigma:=\frac{1}{2}(\theta_1 + \theta_2),
\end{equation}
which is well defined wherever $\rho_1$ and $\rho_2$ are both nonzero. Since the aim is to describe the system in regions where the condensates are close to their (nonzero) ground state values, this restriction is not problematic. Note that the magnetic field $B_k = \varepsilon_{ijk}\partial_i p_j$. This gives us the minimal set of gauge invariant quantities $(\rho_\alpha, \theta_\Delta, p_i)$ where
\begin{equation}
\theta_\Delta:=\frac12(\theta_1-\theta_2).
\end{equation}
The condensates may be conveniently expressed
\begin{equation}
\psi_\alpha = \rho_\alpha e^{i(\theta_\Sigma+d_\alpha\theta_\Delta)},
\end{equation}
at the cost of defining the coefficients $d_\alpha=(-1)^{\alpha+1}$.

Localization of magnetic fields and characteristic length scales, can typically be assessed by linearizing the theory around the ground state. To that end, one assumes that, far from any defect, the gauge invariant quantities decay to one of the possible ground state values $(\rho_\alpha, \theta_{\Delta},p_i) \to (u_\alpha, \theta_0, 0)$.  Note that $\theta_0 =0$ or $\pi/2$ in the phase (anti)locked case and $\theta_0 = \pm \pi/4$ for $s+is$, $s+id$ and $p+ip$ materials, which break time reversal symmetry.  This is because we have defined $\theta_\Delta$ to be {\em half} the phase difference $\theta_{12}$. Defining the quantities,
\begin{equation}
\varepsilon_\alpha:=\rho_\alpha-u_\alpha,\qquad \vartheta:=\theta_\Delta-\theta_0,
\end{equation}
the system is close to the chosen ground state precisely when $\varepsilon_\alpha$, $\vartheta$ and $p_i$ are small. As these are small, we then assume that only linear terms contribute to the field equations, which we may derive by expanding the free energy up to quadratic terms in $(\varepsilon_\alpha,\vartheta,p_i)$ and considering its variation. It will be convenient to define the matrices,
\begin{equation}
\Q^{\alpha\beta}_{ij} := Q^{\alpha\beta}_{ij} \exp i \left( d_\beta - d_\alpha \right) \theta_0,
\label{eq:Kmat}
\end{equation}
which enjoy the same symmetry as the anisotropy matrices: $\Q^{\alpha \beta}_{ij} = \overline{\Q^{\beta\alpha}_{ji}}$.  Note that $\Q^{11}=Q^{11}$, $\Q^{22}=Q^{22}$, $\Q^{12}=e^{-2i\theta_0}Q^{12}$ and $\Q^{21}=e^{2i\theta_0}Q^{21}$, so passing from $Q$ to $\Q$ amounts to twisting the off-diagonal matrices by the ground state value of the phase difference. With this notation, the linearized free energy density  is
\begin{widetext}
\begin{eqnarray}
\nonumber \mathcal{E}_{lin} &=& \frac{1}{2} \Q^{\alpha\beta}_{ij}(\partial_i \varepsilon_\alpha + i u_\alpha (p_i - d_\alpha \partial_i \vartheta) )(\partial_j \varepsilon_\beta - i u_\beta (p_i - d_\beta \partial_i \vartheta))\\
& & + \frac{1}{4} (\partial_i p_j - \partial_j p_i) (\partial_i p_j - \partial_j p_i) + \frac{1}{2} \mathcal{H}_{\alpha\beta} \varepsilon_\alpha \varepsilon_\beta + \mathcal{H}_{\alpha 3} \varepsilon_\alpha \vartheta + \frac{1}{2} \mathcal{H}_{33} \vartheta^2,
\end{eqnarray}
\end{widetext}
where $\mathcal{H}_{ab}$ is the $3\times 3$ Hessian matrix of second partial derivatives of $F_P$ with respect to the variables $(\rho_1,\rho_2,\theta_\Delta)$ evaluated at the chosen ground state, $(u_1,u_2,\theta_0)$.
This leads to the linear equations of motion,
\begin{eqnarray}
\nonumber -\reK^{\alpha\beta}_{ij} \partial_i \partial_j \varepsilon_\beta - \imK^{\alpha\beta}_{ij} u_\beta (\partial_i p_j - d_\beta \partial_i \partial_j \vartheta )& &\\ + \mathcal{H}_{\alpha\beta} \varepsilon_\beta + \mathcal{H}_{\alpha 3} \vartheta =0 \label{eq:eom_eps}\\
\nonumber - \reK^{\alpha\beta}_{ij} u_\alpha u_\beta d_\alpha (d_\beta \partial_i \partial_j \vartheta - \partial_i p_j) \\+ \imK^{\alpha\beta}_{ij} u_\beta d_\beta \partial_i \partial_j \varepsilon_\alpha + \mathcal{H}_{3\alpha} \varepsilon_\alpha + \mathcal{H}_{33} \vartheta = 0 \label{eq:eom_phase}\\
\nonumber -\partial_j^2 p_i + \partial_i \partial_j p_j - \imK^{\alpha\beta}_{ij} u_\alpha \partial_j \varepsilon_\beta \\+ \reK^{\alpha\beta}_{ij} u_\alpha u_\beta (p_j - d_\beta \partial_j \vartheta) =0, \label{eq:eom_p}
\end{eqnarray}
where $\reK$ and $\imK$ denote the real and imaginary parts of $\Q$.
From \eqref{eq:eom_p}, or by direct calculation, we may deduce that the total supercurrent, to linear order in small quantities, is
\begin{equation}
J_i = \imK^{\alpha\beta}_{ij} u_\alpha \partial_j \varepsilon_\beta - \reK^{\alpha\beta}_{ij} u_\alpha u_\beta (p_j - d_\beta \partial_j \vartheta ).
\end{equation}
We note that the coupling of the equations depends critically on whether $\imK$ is nonzero, and that this may happen even if the original $Q$ matrices are purely real if the ground state has complex phase difference (meaning $\theta_{12}\neq 0,\pi$).  

The linearized field equations are, in general, anisotropic, so the
length scales describing decay from a localized defect to the ground state depend on the spatial direction along which decay occurs. To analyze this, we choose and fix a direction $\nvec$ in physical space and then impose on \eqref{eq:eom_eps}, \eqref{eq:eom_phase}, \eqref{eq:eom_p} the ansatz that $\eps_\alpha$, $\vartheta$ and $p_i$ are translation invariant orthogonal to $\nvec$. In practice, the most convenient way to implement this ansatz is to rotate to a new coordinate system $(x_1,x_2,x_3)$, such that the $x_1$ axis is aligned with our chosen direction $\nvec$. We then seek solutions which are independent of $(x_2,x_3)$. 

This amounts to choosing an $SO(3)$ matrix $R$ whose columns are the chosen orthonormal basis, the first of which is $\nvec$ and then transforming the $Q$ matrices according to the rule
\begin{equation}\label{trid}
Q^{\alpha\beta}\mapsto R^TQ^{\alpha\beta}R.
\end{equation}
Note that the phase-twisted anisotropy matrices $\Q^{\alpha\beta}$ and their real and imaginary parts $\reK^{\alpha\beta},\imK^{\alpha\beta}$ also transform in the same way. 

Having rotated our coordinate system and imposed the ansatz that $\eps_\alpha$, $\vartheta$ and $p_i$ depend only on $x_1$, the linearized field equations \eqref{eq:eom_eps}, \eqref{eq:eom_phase}, \eqref{eq:eom_p} reduce to a coupled linear system of ordinary differential equations for 
\begin{equation}
\vec{w}(x_1)=(\varepsilon_1(x_1),\varepsilon_2(x_1),\vartheta(x_1),p_1(x_1),p_2(x_1),p_3(x_1))
\end{equation}
where we have written the gauge invariant vector field $p_i$ in the new basis. The resulting coupled linear system may be economically written,
\begin{equation}
\AA \frac{d^2\vec{w}}{dx_1^2} + \BB \frac{d\vec{w}}{dx_1} + \CC\vec{w} = 0,
\label{Eq:polynomial}
\end{equation}
where $\AA, \BB, \CC$ are the real $6 \times 6$ matrices.
\begin{eqnarray}
\AA &=& \left( \begin{array}{cc} a & 0\\ 0 & a' \end{array} \right),\\
a &:=& \left( \begin{array}{ccc} -\reK^{11}_{11} & -\reK^{12}_{11} & \imK^{1\beta}_{11} u_\beta d_\beta \\ -\reK^{21}_{11} & -\reK^{22}_{11} & \imK^{2\beta}_{11}u_\beta d_\beta \\ \imK^{1\beta}_{11} u_\beta d_\beta & \imK^{2\beta}_{11} u_\beta d_\beta & -\reK^{\alpha\beta}_{11}u_\alpha u_\beta d_\alpha d_\beta \end{array}\right),\\
a' &:=& {\rm diag}(0,-1,-1,-1),\\
\BB &=& \left( \begin{array}{cc} 0 & b \\ -b^T & 0 \end{array} \right), \\
b &:=& \left( \begin{array}{ccc} -\imK^{1\beta}_{11} u_\beta & -\imK^{1 \beta}_{12} u_\beta & -\imK^{1\beta}_{13} u_\beta \\ -\imK^{2\beta}_{11}u_\beta & - \imK^{2\beta}_{12} u_\beta & - \imK^{2\beta}_{13} u_\beta \\ \reK^{\alpha\beta}_{11} u_\alpha u_\beta d_\alpha & \reK^{\alpha\beta}_{12} u_\alpha u_\beta d_\alpha & \reK^{\alpha\beta}_{13} u_\alpha u_\beta d_\alpha \end{array} \right),\quad \\\CC &=& \left( \begin{array}{cc} \mathcal{H} & 0 \\ 0 & \left< \reK \right> \end{array} \right)\\
\left< \reK \right>_{ij} &:=& u_\alpha \reK^{\alpha\beta}_{ij} u_\beta.
\end{eqnarray}

Note that $\AA$ and $\CC$ are symmetric while $\BB$ is skew, and that all the matrices depend implicitly on the chosen direction $\nvec$ through the transformation \eqref{trid}.

The linearised system of field equations \eqref{Eq:polynomial} describes how a system recovers from a perturbation in the $\nv$-direction, under the assumption of translation invariance orthogonal to $\nv$ (for example, how the system behaves near the boundary of a superconductor with normal $\nv$, subject to an external magnetic field).
We seek solutions of the form
\begin{equation}\label{klepto}
\vec{w}(x_1)=\vec{v}e^{-\mu x_1}
\end{equation}
where $\vec{v}$ is a constant vector and ${\rm Re}\mu>0$, so that all fields decay to their ground state values as $x_1\to\infty$. We interpret $\vec{v}$ as a normal mode of the system about the chosen ground state, $\mu$ as the associated field mass, and $\lambda=1/\mu$ as the associated length scale. Given such a solution, let $\vec{z}=-\mu\vec{v}$. Then $(\vec{v},\vec{z})$ satisfies  the linear system
\begin{equation}
\Omega\left(\begin{array}{c} \vec{v}\\ \vec{z}\end{array}\right)
=\frac1\mu\left(\begin{array}{c} \vec{v}\\ \vec{z}\end{array}\right)
\end{equation}
where $\Omega$ is the $12\times12$ matrix
\begin{equation}
\Omega:= \left(\begin{array}{cc} \CC^{-1} \BB & \CC^{-1} \AA \\ -I_6 & 0 \end{array} \right).
\end{equation}
Hence, $\lambda=1/\mu$ is an eigenvalue of $\Omega$. Conversely, given an
eigenvector $(\vec{v},\vec{z})$ of $\Omega$ corresponding to a nonzero  eigenvalue $1/\mu$, $\vec{z}=-\mu\vec{z}$ and \eqref{klepto} is a solution of \eqref{Eq:polynomial}. 

We conclude, therefore, that the length scales associated with decay to the ground state in the fixed direction $\nvec$ are those eigenvalues of $\Omega(\nvec)$ with positive real part. Such eigenvalues are solutions of the degree 12 polynomial equation
\begin{equation}
\det\left(\AA - \lambda \BB + \lambda^2\CC\right) = 0.
\label{Eq:kernel}
\end{equation}
It follows from the symmetry properties of $\AA,\BB,\CC$
that \eqref{Eq:kernel} is actually a real degree 6 polynomial equation in $\lambda^2$, so if $\lambda$ is a solution, so are
$-\lambda,\ol\lambda$ and $-\ol\lambda$. Note that $0$ is an eigenvalue of $\Omega$ of algebraic multiplicity $2$ with eigenvector $(0,\ldots,0,1,0,0)$. This should be discarded as it does not correspond to a solution of \eqref{Eq:polynomial}. Of the remaining 10 eigenvalues, precisely 5 have positive real part: these are the 5 length scales we seek. Let us order them by decreasing real part $\lambda_1,\lambda_2,\ldots,\lambda_5$. We call $\vec{v}_1$, the mode corresponding to the longest length scale $\lambda_1$, the {\em dominant mode} since, generically, at large $x_1$, this will dominate the solution of \eqref{Eq:polynomial}. Depending on the details of the defect being studied, it may be, however, that the dominant mode is unexcited, so subleading modes $\vec{v}_2,\vec{v_3},\ldots$ may still be phenomenologically important.

It is important to note that we have not followed the standard simplified approach to dimensional reduction; we have retained all three components for $p_i$. The standard approach, in contrast, assumes that any local magnetic field always occurs in a single direction, with a single attributed length scale, requiring the retention of only a single component for $p_i$. This is only valid if the magnetic modes are entirely decoupled from the matter modes. If they are coupled, spontaneous magnetic field can be excited in any coupled direction, due to excitations in the matter fields. This can cause the excitation of magnetic where one might not expect it, or a change in the local field direction. As we see from the linearized field equations, generically the anisotropy couples all fields together and we must retain all components of $p_i$, and hence the magnetic field. If we neglect any of these components, our ansatz becomes incompatible with the field equations.

In general, the masses $\mu_i=1/\lambda_i$ associated with the mixed modes $\vec{v}_i$ are complex. This causes the fields at large $x_1$ to behave differently from the standard monotonic Meissner effect. Instead, the fields will exhibit oscillatory behaviour as they decay, with a frequency determined by the imaginary part of $\mu_i$. For all parameter sets studied in this paper, the imaginary part of $\mu_i$ gave periods much larger than the length scales of the modes. Hence, any oscillatory behaviour for $s+is$ or $s+id$ states should be heavily damped and unobservable in experiment for the parameters we considered. However, note that oscillatory linear modes are observable in $p+ip$ systems \cite{speight2019chiral}.

\subsection{Mixed modes}
In an isotropic multi-component superconductor, the normal modes $v_i$ are separated into matter modes: those associated with the coherence length (linear combinations of the modulus of the order parameters \cite{Babaev2010,Carlstroem2011}) as well as the phase difference (Leggett) mode; and magnetic modes: those associated with the magnetic penetration depth. Our analysis reproduces these separate real length scales (coherence length and magnetic penetration depth) in the isotropic limit $Q^{\alpha\beta}_{ij}=\delta_{\alpha\beta}\delta_{ij}$.

Away from the isotropic limit, and in particular for the case of $s+is$ and $s+id$ superconductors, the normal modes are associated with linear combinations of magnetic and matter degrees of freedom. Hence, we should consider all excitations of our system in terms of these \emph{mixed modes} $\vec{v}_i$ and their corresponding length scales $\lambda_i$, as familiar quantities such as the London penetration length do \emph{not} exist. This leads to an important physical consequence; a general excitation decays with coupled modes, inducing spontaneous magnetic fields.

By spontaneous magnetic fields, we mean emerging   local non-zero magnetic field, despite no matching applied external field. Hence, if there is no applied field to the material, a defect or domain wall will still exhibit local magnetic field. Alternatively, if we apply an external field, such as in the Meissner state, the linearisation still predicts local magnetic field orthogonal to the applied field direction (which is not excited by the applied field itself).

In addition to domain walls and defects, if we apply an external field $H$, such as for the Meissner state, the spontaneous fields will cause magnetic field twisting. If the magnetic component of a coupled mode is not parallel to $H$, the induced magnetic field will twist the local magnetic field away from the direction of $H$.  Hence, in general we would expect the local magnetic field induced by the Meissner state to twist its direction as it decays into the bulk of the superconductor.

It is useful to have a measure of how mixed a given mode is. We can achieve this by considering a general mode as a vector in a 5-dimensional space. Note that while the modes are 6-dimensional, $v^4_i$ is redundant (it does not contribute to either the magnetic field or the condensates) and will be excluded for this discussion.  We define the quantity $\theta^i_m$ as the mixing angle of the $i$th mode,
\begin{equation}
\cos \theta^i_m = \underbrace{\sqrt{|v^1_i|^2 + |v^2_i|^2 + |v^3_i|^2}}_{\text{matter modes}}, \quad \sin \theta^i_m = \underbrace{\sqrt{|v^5_i|^2 + |v^6_i|^2}}_{\text{magnetic modes}}.
\label{Eq:mixing}
\end{equation}
Conceptually, the mixing angle is then the angle that the 5-dimensional vector makes with the region in this space representing pure matter modes.
This allows us to classify each mode as either purely matter ($\theta^i_m = 0$), purely magnetic ($\theta^i_m = \pi/2$) or mixed ($0<\theta^i_m<\pi/2$). The angle can be used as a numerical measure of the strength of the mixing.
When a mode exhibits a large density component and small magnetic component, we call such a model density-dominated or vice versa. 

\subsection{Long range dominant modes}

To understand a given field at long range, we must first consider the leading mode $\vec{v}_1$. If this mode is excited by the excitation ($c_1 \neq 0$), then $\vec{v}_1$ is the dominant eigenvector for that field and the long-range behaviour is described by that mode. However, if the mode is not excited, then we must consider the next mode $\vec{v}_2$ and so on. Hence the dominant mode at long range will be the first excited mode.

Previous work has assumed that $B_{2}=B_{1}=0$, forcing any spontaneous magnetic field to be in the $x_3$-direction. However in our case, we consider the more general situation, where the magnetic field is everywhere orthogonal to $x_1$ but is not assumed to lie in a fixed direction in the $(x_2,x_3)$ plane. Hence, for a given linear solution, the magnetic field in our chosen orthonormal basis is $B_{lin}(x_1) = (0,-p_3',p_2')$, so the direction of spontaneous magnetic field for a given mode can be approximated as,
\begin{equation}
{B}_{lin} \parallel {\rm Re}( 0, v_i^6, -v_i^5 ).
\label{eq:linBdir}
\end{equation}

The dominant eigenvalue with magnetic component will determine the direction of the magnetic field at long range. If this does not match the magnetic field direction for the nonlinear part of the defect (for example the spontaneous magnetic field for a domain wall, or the direction of external field for a Meissner state), then the magnetic field will exhibit twisting as the fields decay spatially from $x_1=0$ (nonlinear dominated) to $x_1 \rightarrow \infty$ (linear dominated).  This will be most obvious for the Meissner state, where the magnetic field direction can be fixed to be any orthogonal direction on the boundary of the system, allowing up to $\pi/2$ twisting to occur. 

\subsection{Summary and results}
\begin{figure}
 \centerline{\includegraphics[width=0.9\linewidth,trim={0.1cm 0.1cm 0.1cm 0.1cm},clip]{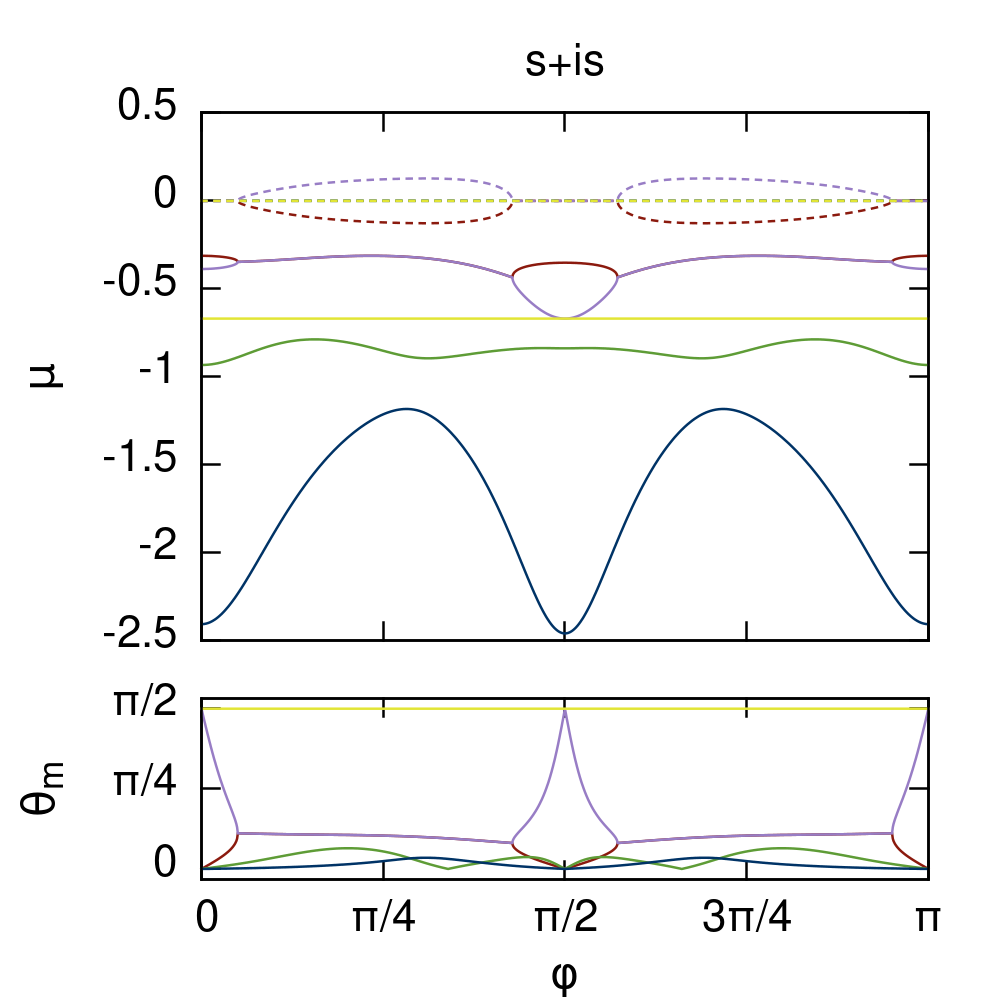}}
 \caption{Plot of the linear mode mixing $\theta_m$ and masses $\mu$ for an $s+is$ superconductor with parameters given in the appendix. The direction of the field variation is $\hat{x}_1 = (\cos \varphi, 0 , \sin \varphi)$,  where $\varphi = 0$ corresponds to the basal plane.  Note that an $s+is$ superconductor is $SO(2)$ symmetric about the $z$-axis. The top panel plots the masses (inverse length scales) $\mu_i = \frac{1}{\lambda_i}$. Each $\mu_i$ is a different colour, with the real part plotted as a solid line and the imaginary part as a dashed line of the same colour. The bottom panel plots the mixing angle $\theta^i_m$ of each mode, where the colours of the modes match the colours of the corresponding mass above (note $\theta_m = 0$ is a pure matter mode and $\theta_m = \pi/2$ is a pure magnetic mode).It can be seen that the linear modes for an $s+is$ system decouple when the fields vary in the basal plane or the $z$-axis direction, but non-trivially couple when they do not.}
 \label{Fig:ls_sis}
\end{figure}

The solutions to the linear equations above, for the parameters given in the appendix, are plotted in \figref{Fig:ls_sid} for $s+id$ and \figref{Fig:ls_sis} for $s+is$.  In both figures we generally see significant mixing, dependent on the orientation of $\nvec$ (the direction along which the fields vary). For both $s+is $ and $s+id$, we observe that when $\nvec$ corresponds to a crystal axis ($\nvec = \hat{x}$, $\hat{y}$ or $\hat{z}$), all mixing disappears. This suggests that excitations with fields that vary solely in the direction of a crystal axis, will exhibit no spontaneous magnetic fields.

If we consider some specific values of $\nvec = (\cos \omega \sin \varphi, \sin \omega \cos \varphi, \sin \varphi)$ for $s+id$ superconductors, we can understand what the linearization predicts in detail. For example, consider $\varphi = 0$, $\omega = \pi/4$ leading to the linear solution, 
\begin{eqnarray}
\mu_1 = 0.33, & \quad v_1 = \left( 0, 0, 0.91, 0.27, 0, 0 \right)^T, \nonumber \\
\mu_2 = 0.39, & v_2 = \left( 0, 0, 0, 0, 0, 1 \right)^T, \nonumber \\
\mu_3 = 0.65 \pm i 0.084, & v_3 =  \left( \begin{array}{c} 0.343 \mp i 0.12 \\ -0.044 \mp i 0.144 \\ 0 \\ 0 \\  -0.55 \mp i 0.485 \\ 0  \end{array} \right), \nonumber \\
\mu_4 = \ol{\mu_3}, & v_4 = \ol{v_3},\nonumber \\
\mu_5 = 1.63, &\, v_5 = \left( 0.0297, 0.509, 0, 0, -0 .116, 0 \right)^T, \nonumber
\label{eq:linHalf}	
\end{eqnarray}
where we have used our freedom to set $x_3 = z$. There are three mixed modes here $v_3$, $v_4$ and $v_5$, which all couple magnetic field in the ${x}_3 = {z}$ direction with the matter fields. Hence, for a linearly dominated system we would expect spontaneous magnetic field only in the $z$-crystalline axis direction. The leading length scale is the phase difference mode $v_1$ followed by the purely magnetic mode $v_2$. This means if the mode $v_2$ is excited, the magnetic field will twist in the $\hat{x}_2 = (-1/\sqrt{2}, 1/\sqrt{2}, 0)$ direction.

\begin{figure}[h!]
	\centerline{\includegraphics[width=0.9\linewidth]{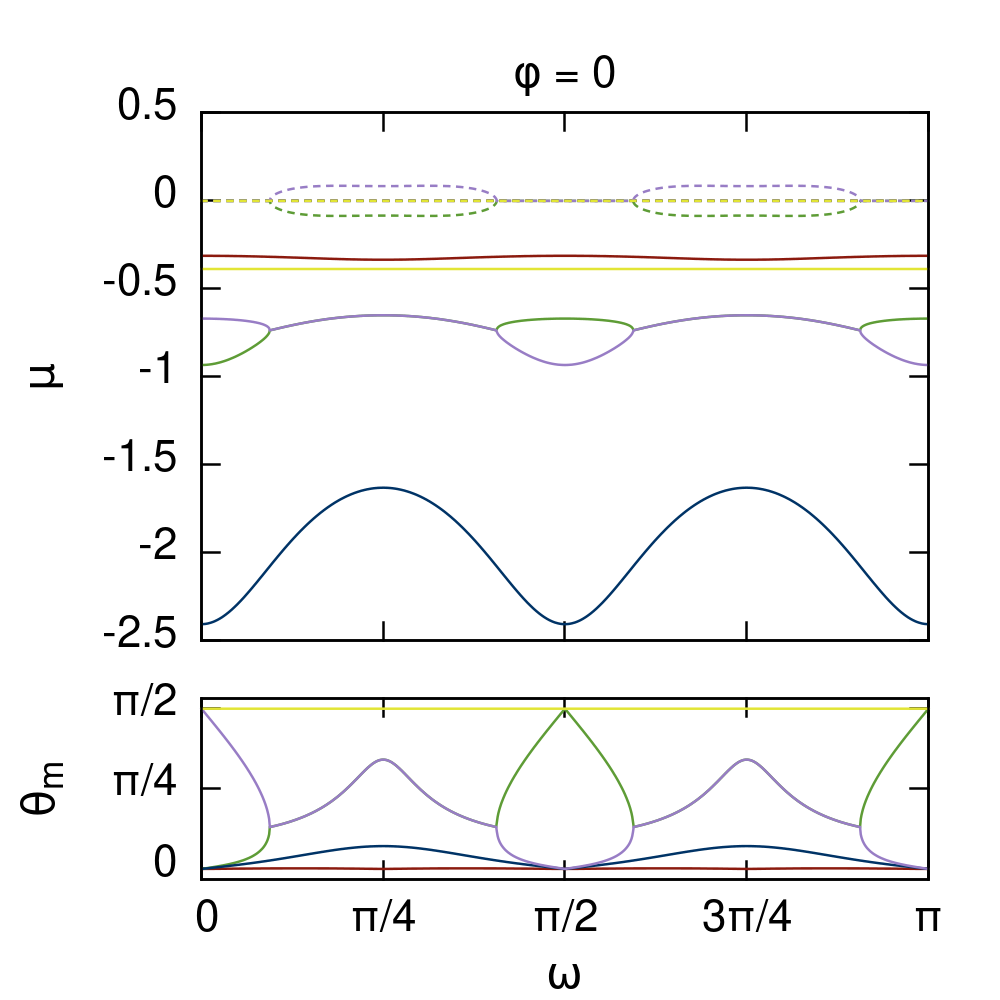}}
	\centerline{\includegraphics[width=0.9\linewidth]{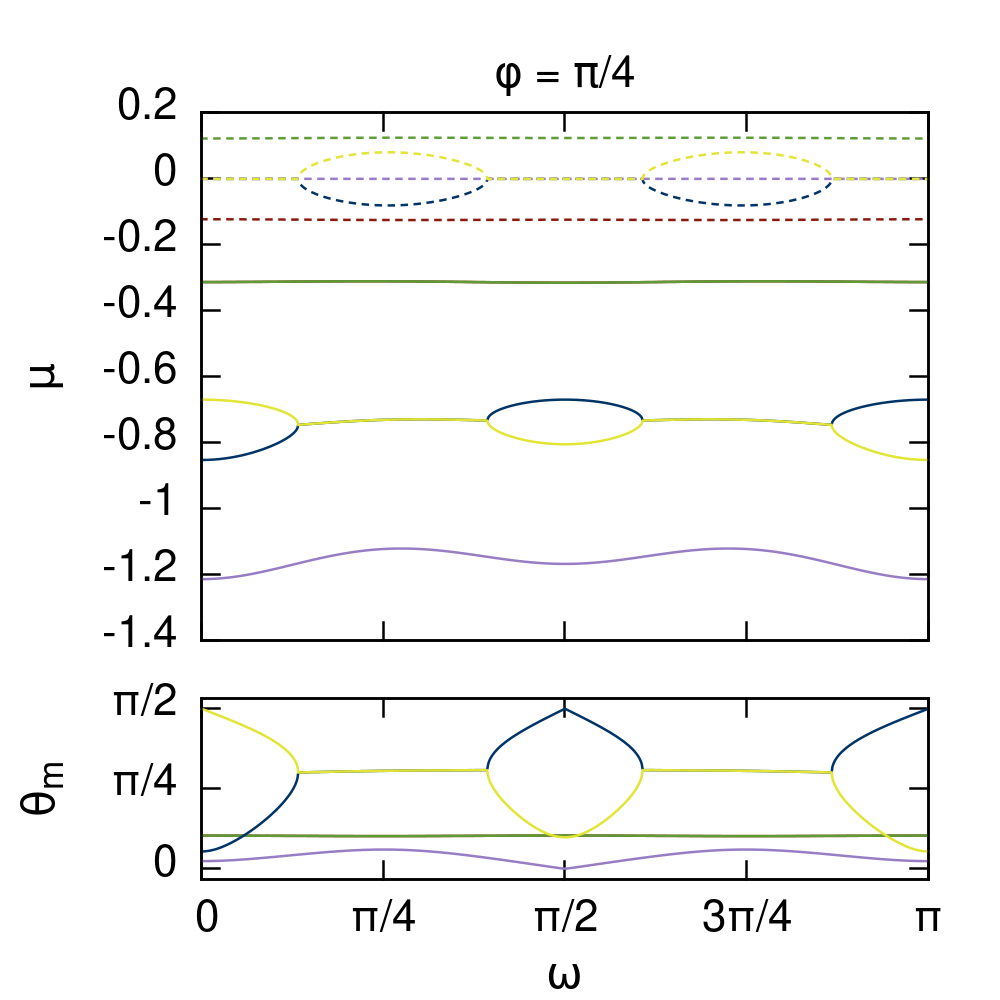}}
	\caption{ Plot of the linear mode mixing $\theta_m$ and masses $\mu$ for an $s+id$ superconductor with parameters given in the appendix. The direction of the field variation is $\hat{x}_1 = (\cos \omega \cos \varphi, \sin \omega \cos \varphi, \sin \varphi)$,  with $\varphi = 0$ (basal plane) for the left plot and $\varphi = \pi/4$ for the right plot.  $\omega$ determines the angle about the $z$-axis. The top panels plot the masses (inverse length scales) $\mu_i = \frac{1}{\lambda_i}$. Each $\mu_i$ is a different colour, with the real part plotted as a solid line and the imaginary part as a dashed line of the same colour. The bottom panel plots the mixing angle $\theta^i_m$ of each mode, where the colours of the modes match the colours of the corresponding mass above (note $\theta_m = 0$ is a pure matter mode and $\theta_m = \pi/2$ is a pure magnetic mode). It can be seen that the linear modes for an $s+id$ system decouple when the fields vary in the direction of a crystalline axis but non-trivially couple when they do not.}
	\label{Fig:ls_sid}
\end{figure}

If we consider the linear solution on any great circle that connects crystalline axes, e.g. $\nvec = (\cos \omega, \sin \omega, 0)$ or $\hat{x}_1 = (\cos \omega, 0, \sin \omega)$ for $\omega \in [0,2\pi]$, the behaviour of the linear modes is similar  to that discussed above.  Hence, they will all exhibit mixing for a single magnetic field direction. Note that, due to the $SO(2)$ symmetry of $s+is$ superconductors,  all orientations can be described by the second of these families and hence exhibit this mixing behaviour.

If we consider a direction for $s+id$ that is not on one of these great circles e.g.  $\varphi = \pi/2$ and $\omega =  \pi/2$, or $\nvec = (1/2, 1/2, 1/\sqrt{2})$, we observe mixing in multiple magnetic field directions.  The linear solution for this orientation has modes corresponding to four different spontaneous magnetic field directions, leading to a complicated spontaneous magnetic field response, with non-trivial magnetic field twisting. However, we can predict that at long range $x_1 \rightarrow \infty$, the leading mode $\mu_1$ will dominate and the magnetic field will twist approximately in the $\hat{x}_2 = (-1/\sqrt{2}, 1/\sqrt{2}, 0)$ direction.

\section{Meissner state}\label{sec:Meissner}

We now consider the effect of applying an external magnetic field to a superconducting material, requiring us to solve the full nonlinear equations of motion in \eqref{Eq:GLeom}. In particular we model a superconductor/insulator boundary as a semi-infinite superconductor $\Omega$ occupying the half-space $x_1 \geq 0$, where $\hat{x}_1$ is the inward pointing normal. An external magnetic field $H = H_0 \hat{x}_3$, orthogonal to the boundary normal ($\hat{x}_1 \cdot \hat{x}_3 = 0$) is applied. This excites the superconducting fields, that decay orthogonally from the boundary into the bulk of the system, dimensionally reducing the problem to a 1-dimensional variational problem on $x_1\in [0,\infty)$.

We first perform a transformation of coordinates from the crystaline basis $(x,y,z)$ to the excitation basis $(x_1,x_2,x_3)$. Note, our new first coordinate is the inward pointing normal and the direction of field variation $\hat{x}_1$; and the third is the external field direction $\hat{x}_3 = \hat{H}$. This coordinate transformation is performed by transforming the anisotropy matrices according to \eqref{trid}.

This allows us to dimensionally reduce the nonlinear field equations to the half-line, by substituting the following ansatz into \eqref{Eq:GLeom},
\begin{align}   \label{Eq:ansatz}
\psi_\alpha &= \psi_\alpha(x_1)\\  \nonumber
A &= A_1(x_1)\hat{x}_1 + A_2(x_1)\hat{x}_2 + A_3(x_1)\hat{x}_3, 
\end{align}
As the fields are dependent on $x_1$ only, the magnetic field has two non-zero components $B = (0,B_2,B_3) = (0,-\partial_1 A_3, \partial_1 A_2)$, both orthogonal to $x_1$. Due to our choice of orthonormal basis, $B_3$ measures the strength of the local magnetic field in the direction of the applied external field and $B_2$ the strength orthogonal to this.

We emphasise that the familiar way of considering a one-dimensional excitation, is to retain only one gauge field component ($A_2$), effectively fixing the magnetic field direction in the applied field direction $\hat{x}_3$. It is clear that this ansatz is {\em not} consistent with the field equations \eqref{Eq:GLeom}
for general choices of anisotropy $Q$. Hence, numerically minimizing $F$ with a single gauge field component will \emph{not} lead to solutions of the full three-dimensional equations of motion. While we can assume the fields have translational symmetry (independent of $x_2$ and $x_3$), we must retain all three gauge field components and hence two orthogonal directions of magnetic field $B_2$ and $B_3$.

By retaining all three components of the gauge field,  we open up the possibility of magnetic field twisting. To measure this, we will consider what we dub the twisting angle,
\begin{equation}
\cos \theta_t = \left(B_3/\sqrt{B_3^2 + B_2^2}\right)
\label{eq:twisting}
\end{equation}

Once translational invariance is applied, we seek global minimisers of the Gibbs free energy of the system,
\begin{equation}
G=\int_\Omega{\cal F}-H_i\int_{\Omega}B_i+\int_{\partial\Omega}{\cal F}_{\mbox{surf}}
\label{Eq:Gibbs}
\end{equation}
subject to natural boundary conditions (detailed in the appendix), where $\mathcal{F}$ is the free energy density. Since we are interested in the bulk behaviour in this paper, we neglect  surface contributions \cite{samoilenka2020microscopic} and set $\mathcal{F}_{\mbox{surf}} = 0$. The external field $H_i$ has no effect on the bulk equations of motion in \ref{Eq:GLeom} and leads to purely boundary effects. The sample is assumed to be infinite in size, with the right hand numerical boundary deep in the bulk, which can be fixed without loss of generality to the ground state,
\begin{equation}
\psi_1 = u_1, \quad \psi_2 = u_2 e^{i\frac{\pi}{2}}, \quad A_i = 0 \ .
\end{equation}

We numerically evolved the system in \eqref{Eq:Gibbs}, using a gradient decent method, where we have discretized the model on a regular one-dimensional grid of $N$ lattice sites with spacing $h > 0$. The plots in this section were simulated with values $N=1001$ and $h = 0.05$. We approximated the 1st and 2nd order spatial derivatives using central 4th order finite difference operators, yielding a discrete approximation $E_{dis}$ to the functional $G(\phi)$, where $\phi = (\psi_\alpha, A_i)$ are the collected fields. Mathematically, this is a function $E_{dis} : \mathcal{C}\rightarrow \mathbb{R}$, where the discretised configuration space is $\mathcal{C} = (\mathbb{C}^2\times \R^3)^N \simeq \mathbb{R}^{7N}$. Hence, we represent the field configuration by a vector $\phi\in\R^{7N}$. To find a local minimum of $E_{dis}$ w.r.t. the collected fields $\phi$, we use an arrested Newton flow algorithm. That is, we solve for the motion of a notional ``particle" in $\mathcal{C}$, with trajectory $\phi(t)$, moving according to Newton's law in the potential $E_{dis}$,
\begin{equation}
\ddot{\phi}_i = - \frac{\partial E_{dis}(\phi)}{\partial \phi_i},
\label{eq:grad}
\end{equation}
starting from rest ($\dot\phi(0)=0$) at an initial configuration 
$\phi(0)\in\mathcal{C}$.
The time evolution is approximated using a simple Euler method. That is, we evolve the configuration from time $t$ to time $t+\delta t$ by the rule
\begin{align}
\phi_i(t + \delta t) &= \phi_i(t) + \delta t \, \dot{\phi}_i(t),\\ \dot{\phi}_i(t + \delta t) &= \dot{\phi}_i(t) - \delta t \, \frac{\partial E_{dis}}{\partial \phi_i}\bigg|_{\phi(t)},
\end{align}
where $\delta t>0$ is a fixed small parameter (typically $\delta t=0.1h$).
Evolving this algorithm initially causes the configuration $\phi(t)$ to roll downhill, that is, to relax towards a local minimum, where
\begin{equation}
\frac{\partial E_{dis}}{\partial \phi_i} = 0.
\label{Eq:tol}
\end{equation}
If the algorithm is left to run without any damping, $\phi(t)$ will overshoot the minimum and oscillate indefinitely, so we implement an arresting criterion: as soon as
\begin{equation}
\frac{d E_{dis}(\phi)}{d t} = \sum_{i=1}^{7N}\frac{\partial E_{dis}(\phi)}{\partial \phi_i}\dot\phi_i > 0
\end{equation}
we set $\dot{\phi}(t)=0$ and restart the flow (from $\phi(t)$). 
This condition can be thought of as the force or acceleration being in the opposite half-plane to the velocity. Another commonly used arresting condition is that energy increases on the current time step: 
$$
E_{dis}(\phi(t+\delta t)) > E_{dis}(\phi(t)).
$$ 
Of course, this condition is equivalent to ours in the continuous time limit ($\delta t\to 0$), and is, perhaps conceptually simpler, but has the (significant) disadvantage that it requires the computation of $E_{dis}$ at each time step. In summary, our time stepping algorithm is 
\begin{align}
\phi_i(t + \delta t) &= \phi_i(t) + \delta t \, \dot{\phi}_i(t),\nonumber\\ \dot{\phi}_i(t + \delta t) &= \left\{ \begin{array}{cc} 0 & \mbox{if }  \frac{\partial E_{dis}}{\partial \phi}\bigg|_{\phi(t)} \cdot \dot\phi(t) > 0,\\ \dot{\phi}_i(t) - \delta t \, \frac{\partial E_{dis}}{\partial \phi_i}\bigg|_{\phi(t)} & \mbox{otherwise.} 
\end{array}\right.
\label{Eq:algorithm}
\end{align}
We continue this time evolution until the condition in \eqref{Eq:tol} is met within a given tolerance,
\begin{equation}
\max_{i\in\{1,2,\ldots,7N\}}\left| \frac{\partial E_{dis}(\phi)}{\partial \phi_i} \right| < \mbox{tol}.
\end{equation}
The results reported below used $\mbox{tol} = 10^{-6}$.

\subsection{Meissner State results}

\begin{figure*}
	$s+id$, $\hat{x}_1 = (1,0,0)$
	\centerline{\includegraphics[width=1.15\linewidth,trim={0.1cm 0.2cm 1.0cm 0.1cm},clip]{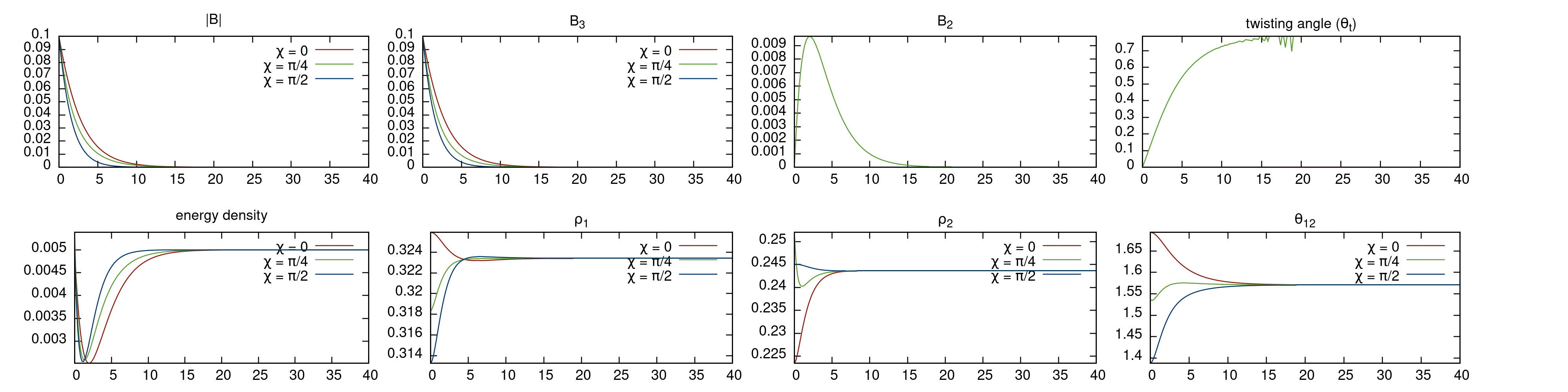}}
	$s+is$, $\hat{x}_1 = (1,0,0)$
	\centerline{\includegraphics[width=1.15\linewidth,trim={0.1cm 0.2cm 1.0cm 0.1cm},clip]{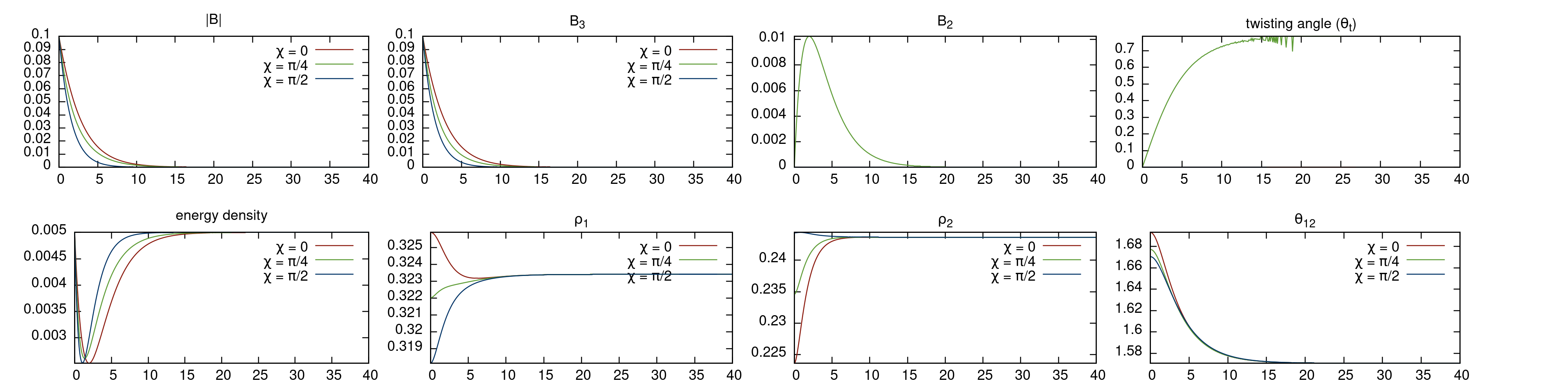}}
	\caption{Meissner state for an $s+id$ (top) and $s+is$ (bottom) system with boundary normal $\hat{x}_1 = (1,0,0)$ and external magnetic field $H = 0.1(0, \cos \chi, \sin \chi)$. The boundary is at $x_1=0$ where $B_3$ measures the strength of local magnetic field in the external magnetic field direction. Comparatively $B_2$ gives the strength of local magnetic field orthogonal to the external field direction, such that $B_3(0) = 0.1$, $B_2(0) = 0$. The twisting angle is given in \eqref{eq:twisting} and determines the amount the local magnetic field twists away from the external field direction. We have also plotted the normalised energy density $\hat{\mathcal{F}} = \mathcal{F} - \mathcal{F}_0$, the condensate densities $\rho_1$,$\rho_2$ and phase difference $\theta_{12} = \theta_1 - \theta_2$. We can see that the magnetic field twists direction as it decays for both $s+is$ and $s+id$, when the applied magnetic field is not in a crystalline-axis direction.}	\label{Fig:Meissner1}
\end{figure*}

\begin{figure*}
	$s+id$, $\hat{x}_1 = (1/\sqrt{2},1/\sqrt{2},0)$
	\centerline{\includegraphics[width=1.15\linewidth,trim={0.1cm 0.2cm 1.0cm 0.1cm},clip]{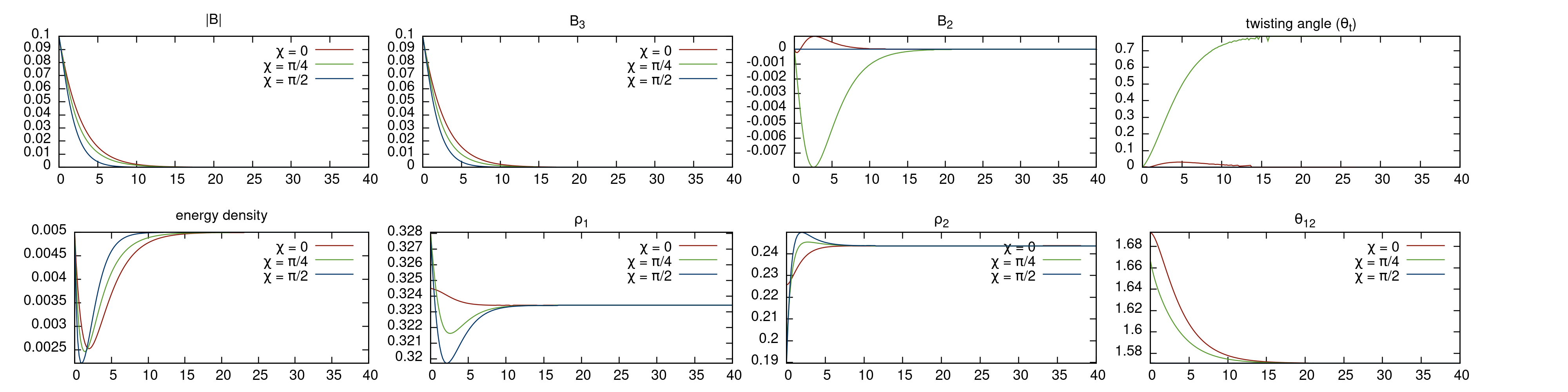}}
	$s+is$, $\hat{x}_1 = (1/\sqrt{2},1/\sqrt{2},0)$
	\centerline{\includegraphics[width=1.15\linewidth,trim={0.1cm 0.2cm 1.0cm 0.1cm},clip]{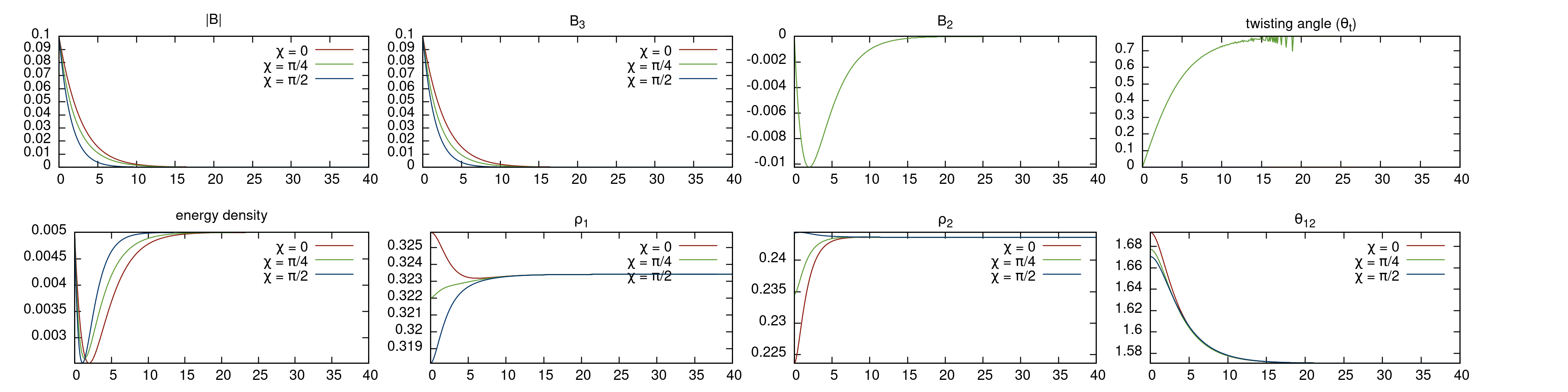}}
	\caption{Meissner state for an $s+id$ (top) and $s+is$ (bottom) system with boundary normal $\hat{x}_1 = (1/\sqrt{2},1/\sqrt{2},0)$ and external magnetic field $H = 0.1(\cos \chi /\sqrt{2}, -\cos \chi /\sqrt{2}, \sin \chi)$. The boundary is at $x_1=0$ where $B_3$ measures the strength of local magnetic field in the external magnetic field direction. Comparatively $B_2$ gives the strength of local magnetic field orthogonal to the external field direction, such that $B_3(0) = 0.1$, $B_2(0) = 0$. The twisting angle is given in \eqref{eq:twisting} and determines the amount the local magnetic field twists away from the external field direction. We have also plotted the normalised energy density $\hat{\mathcal{F}} = \mathcal{F} - \mathcal{F}_0$, the condensate densities $\rho_1$,$\rho_2$ and phase difference $\theta_{12} = \theta_1 - \theta_2$. We can see that for $s+is$ due to symmetry this is equivalent to \figref{Fig:Meissner1}, where as for $s+id$ we see spontaneous magnetic field for multiple directions, causing twisting.}	\label{Fig:Meissner2}
\end{figure*}

\begin{figure*}
	$s+id$, $\hat{x}_1 = (1/2,1/2,1/\sqrt{2})$
	\centerline{\includegraphics[width=1.15\linewidth,trim={0.1cm 0.2cm 1.0cm 0.1cm},clip]{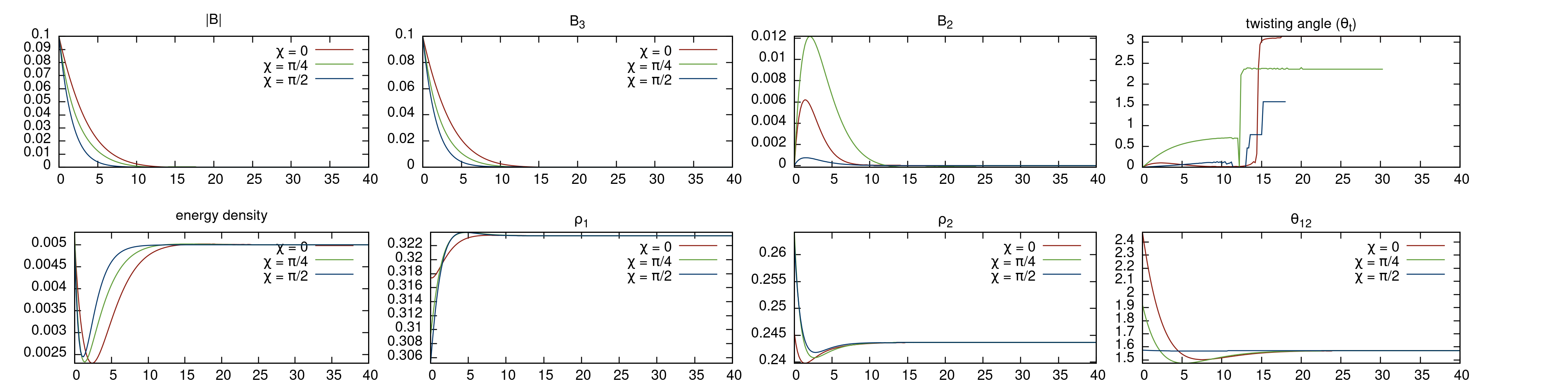}}
	$s+is$, $\hat{x}_1 = (1/2,1/2,1/\sqrt{2})$
	\centerline{\includegraphics[width=1.15\linewidth,trim={0.1cm 0.2cm 1.0cm 0.1cm},clip]{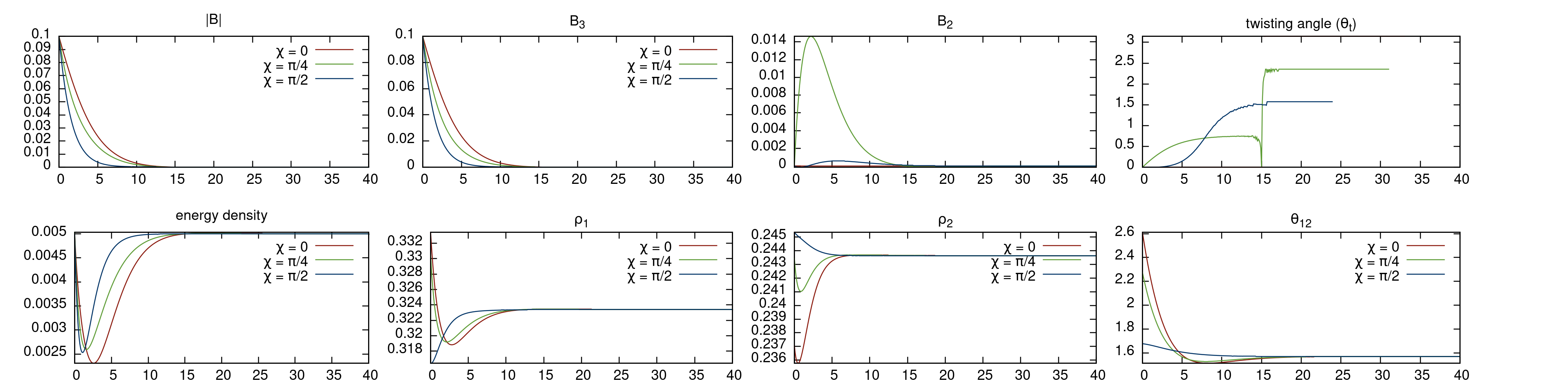}}
	\caption{Meissner state for an $s+id$ (top) and $s+is$ (bottom) system with normal $\hat{x}_1 = (1/2,1/2,1/\sqrt{2})$ and external magnetic field $H = 0.1\cos\chi(1/2,1/2,-1/\sqrt{2}) + 0.1\sin\chi(-1/\sqrt{2},1/\sqrt{2},0)$. The boundary is at $x_1=0$ where $B_3$ measures the strength of local magnetic field in the external magnetic field direction. Comparatively $B_2$ gives the strength of local magnetic field orthogonal to the external field direction, such that $B_3(0) = 0.1$, $B_2(0) = 0$. The twisting angle is given in \eqref{eq:twisting} and determines the amount the local magnetic field twists away from the external field direction. We have also plotted the normalised energy density $\hat{\mathcal{F}} = \mathcal{F} - \mathcal{F}_0$, the condensate densities $\rho_1$,$\rho_2$ and phase difference $\theta_{12} = \theta_1 - \theta_2$. We can see that for $s+is$ and $s+id$ all directions of applied field exhibit spontaneous fields and field twisting.}	\label{Fig:Meissner3}
\end{figure*}

We simulated the boundary problem described above for the parameters given in the Appendix. We simulated multiple orientations of boundary normal $\hat{x}_1$ and applied magnetic field $\hat{x}_3$, uniquely defining the orthonormal basis in \eqref{trid}, with external field strength $H_0 = 0.1$.

In \figref{Fig:Meissner1} the Meissner state with normal $\hat{x}_1 = (1,0,0)$ is plotted, where the applied magnetic field direction is $\hat{x}_3 = (0,\cos \chi, \sin \chi)$ for $\chi = 0, \frac{\pi}{4}, \frac{\pi}{2}$. The linear modes in this direction for an $s+id$ system, shown in \figref{Fig:ls_sid} (at $\omega = 0$), and $s+is$, shown in figure \figref{Fig:ls_sis} (at $\varphi = 0$), predict no mixing of magnetic and matter components. This suggests there is no spontaneous magnetic field in the linear theory for this boundary orientation, regardless of the direction of applied magnetic field. This is also what we observe for the full nonlinear solutions in figure \figref{Fig:Meissner1}, however for $\chi = \pi/4$ we still observe some magnetic field twisting. This is due to both magnetic modes being excited for this orientation (as opposed to one for the other orientations), which decay with different length scales (or masses).

In \figref{Fig:Meissner2} we have plotted the numerical solution with boundary normal $\hat{x}_1 = (1/\sqrt{2}, 1/\sqrt{2}, 0)$ and applied field direction $\hat{x}_3 = (-\cos\chi/\sqrt{2}, \cos\chi/\sqrt{2}, \sin \chi)$. The linear modes for this orientation are given in \eqref{eq:linHalf} and predict spontaneous magnetic field purely in the $\hat{z}$ crystalline axis direction. If this prediction approximates the full nonlinear solutions well, we would expect to observe magnetic field twisting when the applied external field direction $\hat{x}_3$ is orthogonal to the $\hat{z}$-direction but not when it is parallel. This is precisely what we observe, with twisting for $\chi = 0$ but not for $\chi = \pi/2$. In addition, as the leading (purely) magnetic mode is in the  $(1/\sqrt{2},-1/\sqrt{2},0)$ direction, we expect the magnetic field to twist towards this direction as $x_1\rightarrow \infty$ which is what we observe for $\chi = 0,\pi/4$. However, it is expected that this does not occur for $\chi = \pi/2$ as this mode is never excited, due to it being purely magnetic and orthogonal to the applied external field direction $\hat{x}_3$.

Finally, in \figref{Fig:Meissner3} we consider the numerical solution with boundary $\hat{x}_1 = (1/2,1/2,1/\sqrt{2})$ and applied field direction $\hat{x}_3 = \cos \chi (1/\sqrt{2},1/\sqrt{2},0) + \sin \chi (1/2,1/2,-1/\sqrt{2})$. The linear modes for this orientation were observed to have multiple coupled magnetic field directions. This means we would expect spontaneous magnetic field for all choices of external applied field direction, which is what we observe. As all modes are excited, we also expect the magnetic field to twist towards the $(-1/\sqrt{2}, 1/\sqrt{2}, 0)$ direction (corresponding to the leading mode), which is what we observe.

To summarize, the linearization is surprisingly accurate at describing the spontaneous magnetic field response of the full nonlinear Meissner state solutions. The magnetic field twisting is highly dependent on the form of $Q^{\alpha\beta}_{ij}$ and is also significant. This may offer an experimentally viable way of determining the symmetries that a material exhibits when in a superconducting state. 

\section{Domain Walls}
A direct consequence of the $Z_2$ symmetry of $F_p$ in \eqref{eq:Fp} is the existence of domain walls solutions.  These are 1-dimensional excitations that  interpolate between the two distinct, gauge inequivalent ground state values, $\lim_{x_1\rightarrow \pm \infty} \theta_{12} = \mp \pi/2$. The field configurations are independent of all but one spatial coordinate ${x}_1$. In an isotropic two-component BTRS model this forms a 2-dimensional wall in the condensates only,  with normal parallel to $x_1$. However, it has recently been shown that in $s+is$ and $s+id$ models, domain walls also exhibit spontaneous magnetic field \cite{benfenati2020magnetic}. The linearization in section III offers a way of both explaining and predicting the form of these spontaneous fields.  It is important to understand spontaneous fields induced by domain walls (and other defects), as they are important indicators for the underlying pairing symmetries of the host materials. 

We seek one-dimensional solutions to the full nonlinear bulk equations of motion \eqref{Eq:GLeom1} and \eqref{Eq:GLeom}, for both the $s+is$ and $s+id$ models (parameters given in the appendix).  As we are interested in solutions far from any boundary effects, we can fix the boundary conditions such that,
\begin{align} \label{Eq:DW_BCs2}
(\psi_1,\psi_2) &\rightarrow (u_1, -iu_2) , \ &x_1 \rightarrow -\infty \\ \nonumber
(\psi_1,\psi_2) &\rightarrow (u_1, iu_2), \ &x_1 \rightarrow +\infty \\ \nonumber 
(A_1,A_2,A_3) & \rightarrow (0,0,0) , \ & x_1 \rightarrow \pm \infty  \,
\end{align}
where $\hat{x}_1$ is the unit normal of the domain wall. Note, we have transformed from the crystalline basis $(\hat{x},\hat{y},\hat{z})$ to the excitation basis $(\hat{x}_1, \hat{x}_2, \hat{x}_3)$ by transforming the anisotropy matrices according to \eqref{trid}. This leaves all fields dependent on $x_1$ only. In addition,  $A_1 =A_2 =A_3 = 0$ on the boundary is a gauge choice, leading to the finite energy requirement that $\partial_1\psi_1 = \partial_1\psi_2 = 0$ on the boundary.

For a domain wall solution the phase difference $\theta_{12} \in S^1$ interpolates from $\pi/2$ to the antipodal point $-\pi/2$.  This can be achieved by traversing the target $S^1$ clockwise or anticlockwise.  For a BTRS model with no anisotropy, the domain walls corresponding to the different routes are degenerate in energy and have identical forms for the gauge invariant fields $|\psi_\alpha|$.  However, considering these two possible domain wall solutions for a general anisotropic BTRS model, we find that the domain walls are not degenerate in energy. We can see this by considering a simple approximation to a domain wall, allowing only $\theta_{12}$ to depend on $x_1$, while all other quantities are fixed to their ground state values: $\rho_\alpha=u_\alpha$ and $p=0$. Such a configuration has energy (per unit area),
\begin{align}  
\begin{split}
F_{reduced} = \int_{-\infty}^{\infty}  \Bigg\{  \Bigg. & \frac{1}{8} (Q^{11}_{11} u_1^2 + Q^{22}_{11}u_2^2)(\theta_{12}'(x_1))^2 \\
& - \frac{1}{4} Q^{12}_{11} u_1 u_2 \cos{\theta_{12}(x_1)}(\theta_{12}'(x_1))^2 \\
&+ \frac{\eta}{8}u_1^2 u_2^2 \cos{2\theta_{12}} \Bigg. \Bigg\} \ dx_1 .
\end{split} \label{eqn:Fred}
\end{align}
We note that if $Q^{12}_{11} = 0$ then $F_{reduced}$ is invariant under the transformation $\theta_{12} \rightarrow \pi - \theta_{12}$,  which converts between the two domain wall solutions. In addition, as $u_1, u_2 > 0$, when $Q^{12}_{11} \neq 0$ the second term will either be positive definite or negative definite, dependent on the sign of $ Q^{12}_{11} $ and $\cos\theta_{12}$. Hence, if $Q^{12}_{11} > 0 $ then the clockwise domain wall is lower energy and if $Q^{12}_{11} < 0 $ then the anticlockwise domain wall has lower energy. This suggests that the sign of $Q^{12}_{11}$ can be used to predict which of the two domain wall solutions is the global minimiser for a given orientation.  This approximation is rather crude, as it ignores couplings between $\theta_{12}$ and the other fields.  However it seems to capture the behaviour of the systems studied numerically very well. 

\begin{figure*}
	\centering
	\text{\large \boldmath $\hat{x}_1 = (0.1736,0,0.9848)$  }
	\begin{minipage}[c]{.95\textwidth}
		\includegraphics[max size={\textwidth}{\textheight}]
		{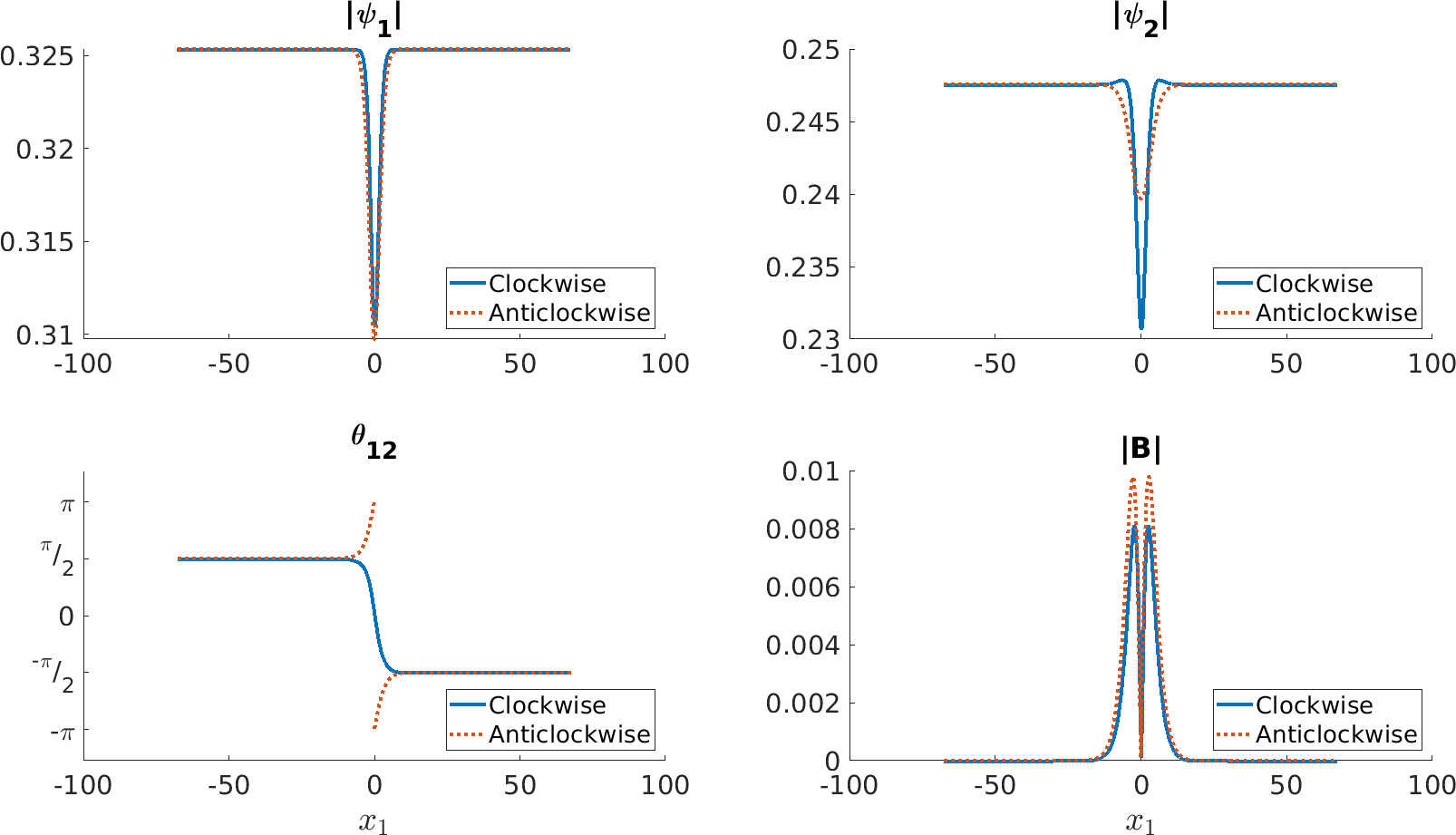}	
		\subcaption{}\label{fig:DWs_plot1}
	\end{minipage}%
	\vfil
	\centering
	\text{\large \boldmath $\hat{x}_1 = (0.3090,-0.9511,0)$  }
	\begin{minipage}[c]{.95\textwidth}
		\includegraphics[max size={\textwidth}{\textheight}]
		{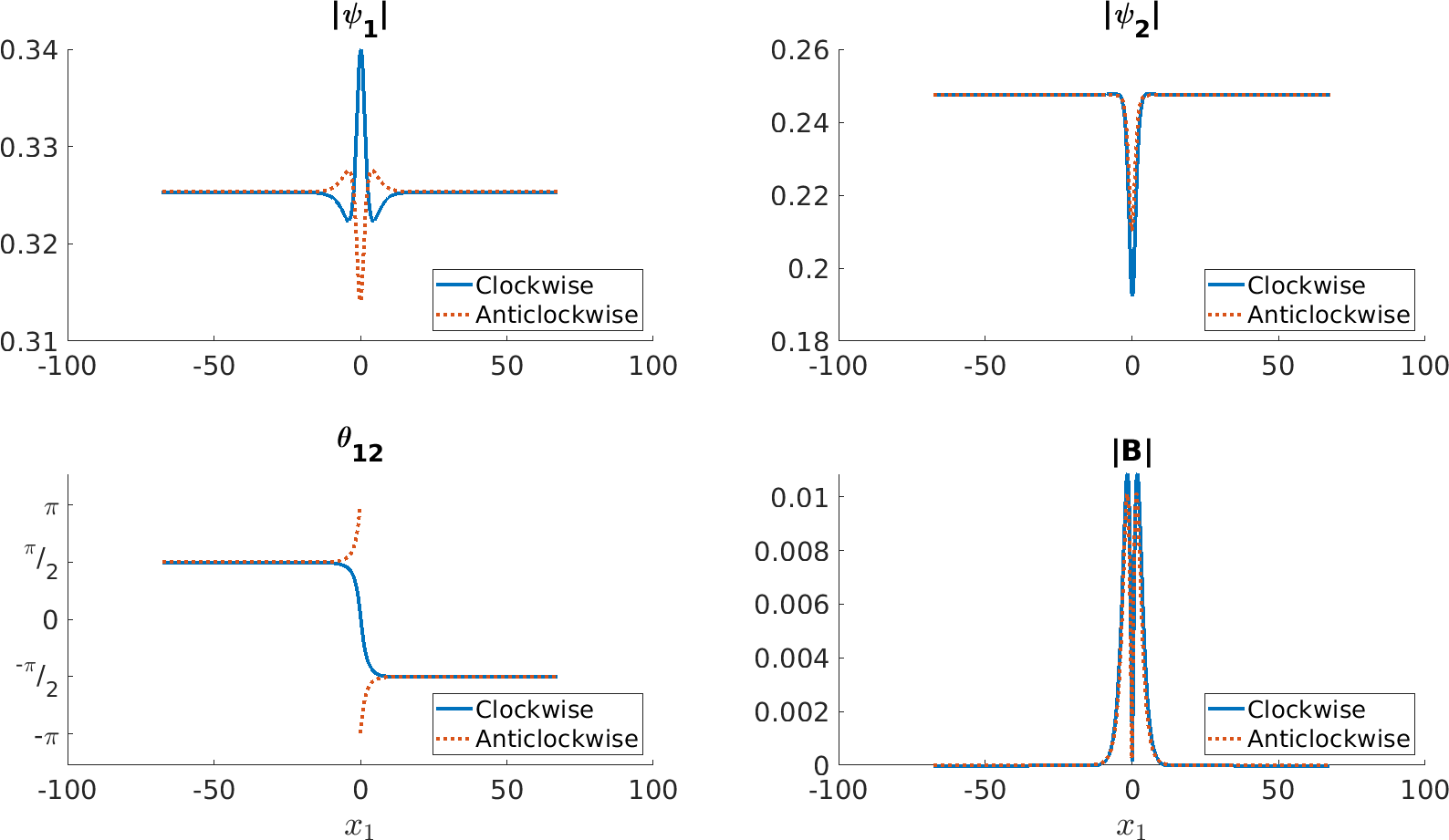}
		\subcaption{}\label{fig:DWs_plot2}
	\end{minipage}
	\caption{Plots of two domain wall solutions corresponding to the phase difference $\theta_{12} = \theta_1 - \theta_2$ winding either clockwise (red) or anticlockwise (blue).  We have plotted the gauge invariant condensate magnitudes $\rho_1$,$\rho_2$ and the total magnetic field strength $|B|$. We can see that the magnetic response of the two different domain walls is different.
	} 
\end{figure*}

We study domain walls by solving the equations of motion in \eqref{Eq:GLeom1} and \eqref{Eq:GLeom} numerically. In particular, we seek 1-dimensional numerical minimizers of the free energy functional in \eqref{Eq:F}. We first choose an orientation (normal) for the domain wall $\hat{x_1}$, which is also the sole spatial dependence for the fields.  We then transform the anisotropy matrices according to \eqref{trid} and dimensionally reduce by assuming that all field derivatives orthogonal to $x_1$ are zero (an effective gauge choice). We then use an arrested Newton flow method (described previously for the Meissner state simulations in \secref{sec:Meissner}), subject to the fixed boundary conditions described in \eqref{Eq:DW_BCs2}. Of course, we now seek to minimize a discrete approximant $E_{dis}$ to the Helmholtz free energy $F$, rather than the Gibbs free nergy $G$, as there is no applied magnetic field. We find numerical minimizers for the parameters described in the appendix, for typical values of $N = 1001$ and $h = 0.15$. 

The initial field configuration $\phi(0)$ was chosen to interpolate the phase difference either clockwise or anti-clockwise,
\begin{equation}
\theta^0_{12}(x) = \left\{\begin{array}{cc} -\frac{\pi}{2} & x < -L\\
-\frac{\pi}{2} \pm \frac{(x+L)\pi}{2L} & |x| \leq L \\
\frac{\pi}{2} & x > L \end{array}\right.
\end{equation}
respectively, where $x = h(i - (N+1)/2)$, $i \in [1,N]$ is the lattice site and the typical width of the initial condition was $2L = 10$. This allows us to consider both the clockwise and anticlockwise domain wall solutions discussed above, chosen by interpolating the phase difference around the target circle in the corresponding direction.

\subsection{Domain Wall Results}

We have plotted examples of both domain wall solutions with normal $\hat{x}_1 = (0.1736,0,0.9848)$ in \figref{fig:DWs_plot1} and $\hat{x}_1 = (0.309, -0.9511, 0)$ in \figref{fig:DWs_plot2}.  Both the clockwise and anticlockwise domain wall solutions exhibit spontaneous magnetic fields for both orientations; however the strengths of the spontaneous fields differ for each solution. This demonstrates that the two domain wall solutions for a given orientation will have distinct experimental signatures. 

\begin{figure*}
	   \centering
	\textbf{\large Total normalised free energy ($\hat{F}$)} \par\medskip\vspace{0.2cm}

	\begin{minipage}[t]{.45\textwidth}
		\includegraphics[max size={\textwidth}{\textheight}]
		{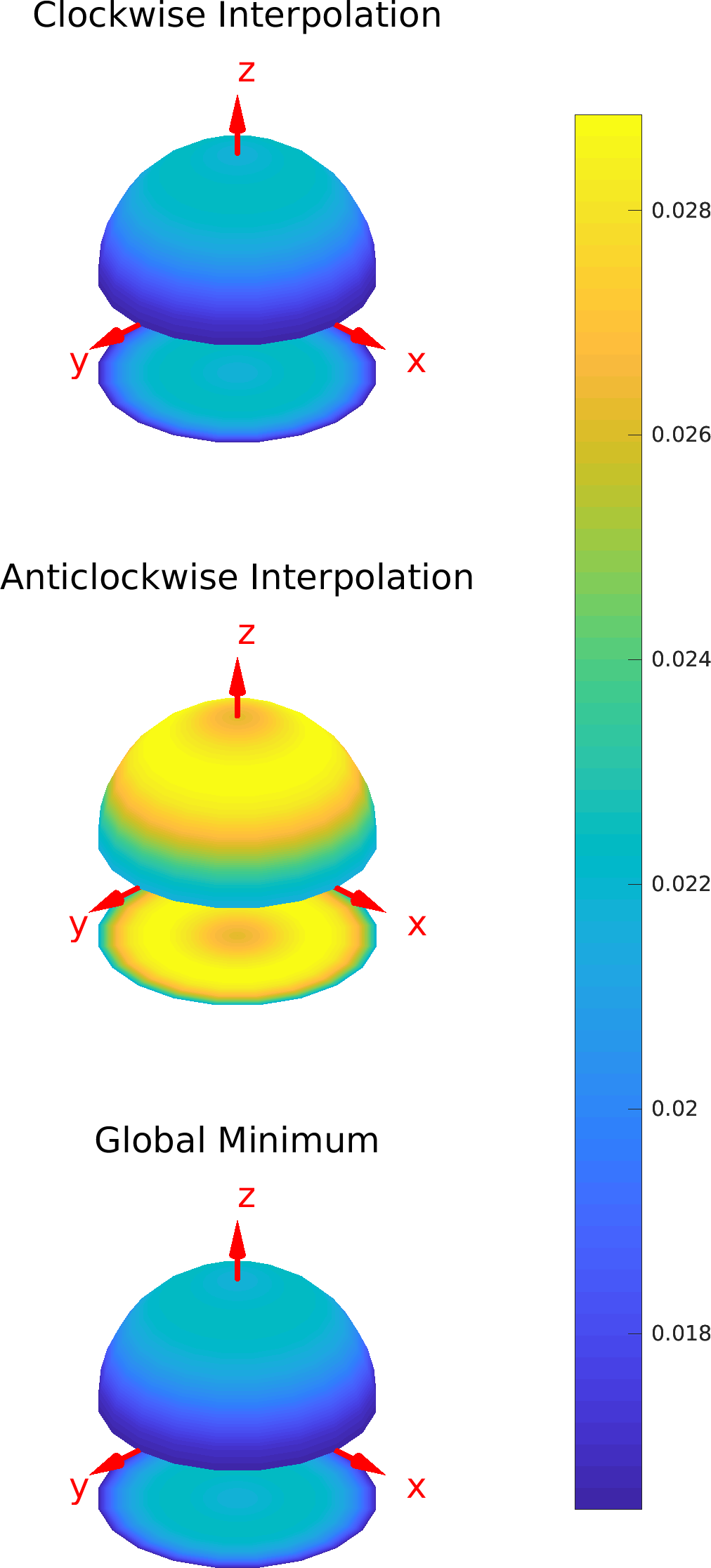}
		\subcaption{s+is} \label{fig:DWs_E_s}
	\end{minipage}%
	\hfill
	\begin{minipage}[t]{.45\textwidth}
		\includegraphics[max size={\textwidth}{\textheight}]
		{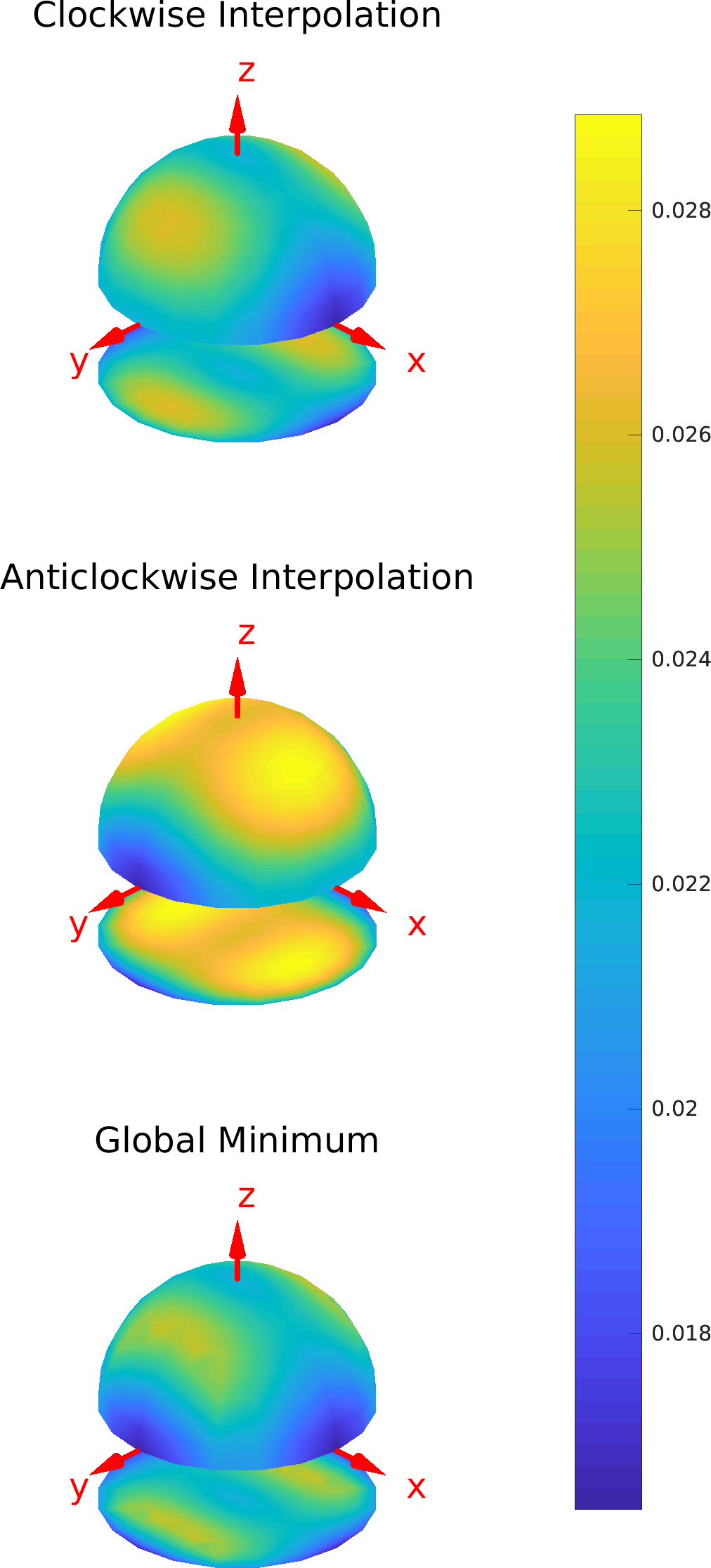}
		\subcaption{s+id}\label{fig:DWs_E_d}
	\end{minipage}
	\caption{Plot of the total normalized free energy $\hat{F} = F - F_0$ of each domain wall solution for all possible orientations.  We have mapped each possible orientation (normal vector) to a point on the unit two-sphere. The sphere has then been coloured by the total normalized free energy of the corresponding domain wall solution. Since both $s+is$ and $s+id$ models are symmetric under $z\mapsto-z$ we plot only the upper hemisphere. There are two non-degenerate domain wall solutions depending on the winding of the phase difference (clockwise or anticlockwise). Note that the minimal energy domain wall is always the clockwise solution for $s+is$ but is orientation dependent for $s+id$.}
\end{figure*}

We have also plotted the total free energy for all possible orientations of the normal $\hat{x}_1$ for an $s+is$ model in  \figref{fig:DWs_E_s} and an $s+id$ model in \figref{fig:DWs_E_d}.  These plots display the free energy for all possible orientations for the normal in the crystalline basis, by mapping each orientation to a point on a unit 2-sphere. Due to the symmetry of $F$ under the reflexion $z\mapsto-z$, it is sufficient to retain only the upper hemisphere of the resulting plot.  Each point is then coloured by the total (normalised) free energy of the numerical solution.

When simulating these sets of solutions, we choose a set of approximately equidistant points on the sphere for $\hat{x}_1$ and use the local minimum from the previous simulation as the initial condition for the next. This preserves whether the domain wall interpolates clockwise or anti-clockwise.

\begin{figure*}
	\centering
	\begin{minipage}[t]{.45\textwidth}
		\centering
		\textbf{\large \boldmath   $\tilde{Q}^{12}_{11} $ }\hspace{2.0 cm} \vspace{0.2cm} \par\medskip
		\includegraphics[max size={\textwidth}{\textheight}]
		{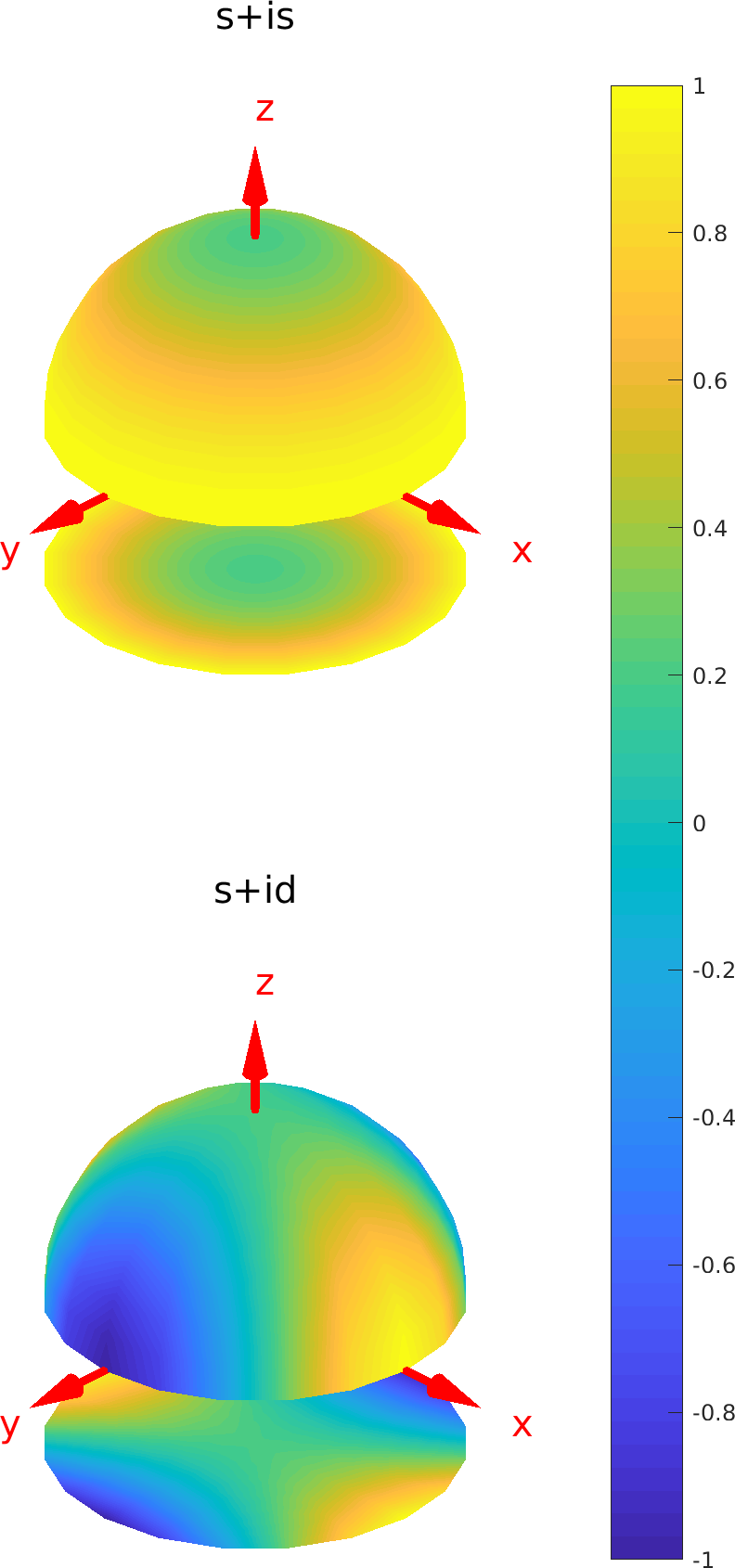}
	\end{minipage}%
	\hfill
	\begin{minipage}[t]{.45\textwidth}
				\centering
		\textbf{\large Free energy difference }\hspace{1.5cm}  \vspace{0.2cm}\par\medskip
		\includegraphics[max size={\textwidth}{\textheight}]
		{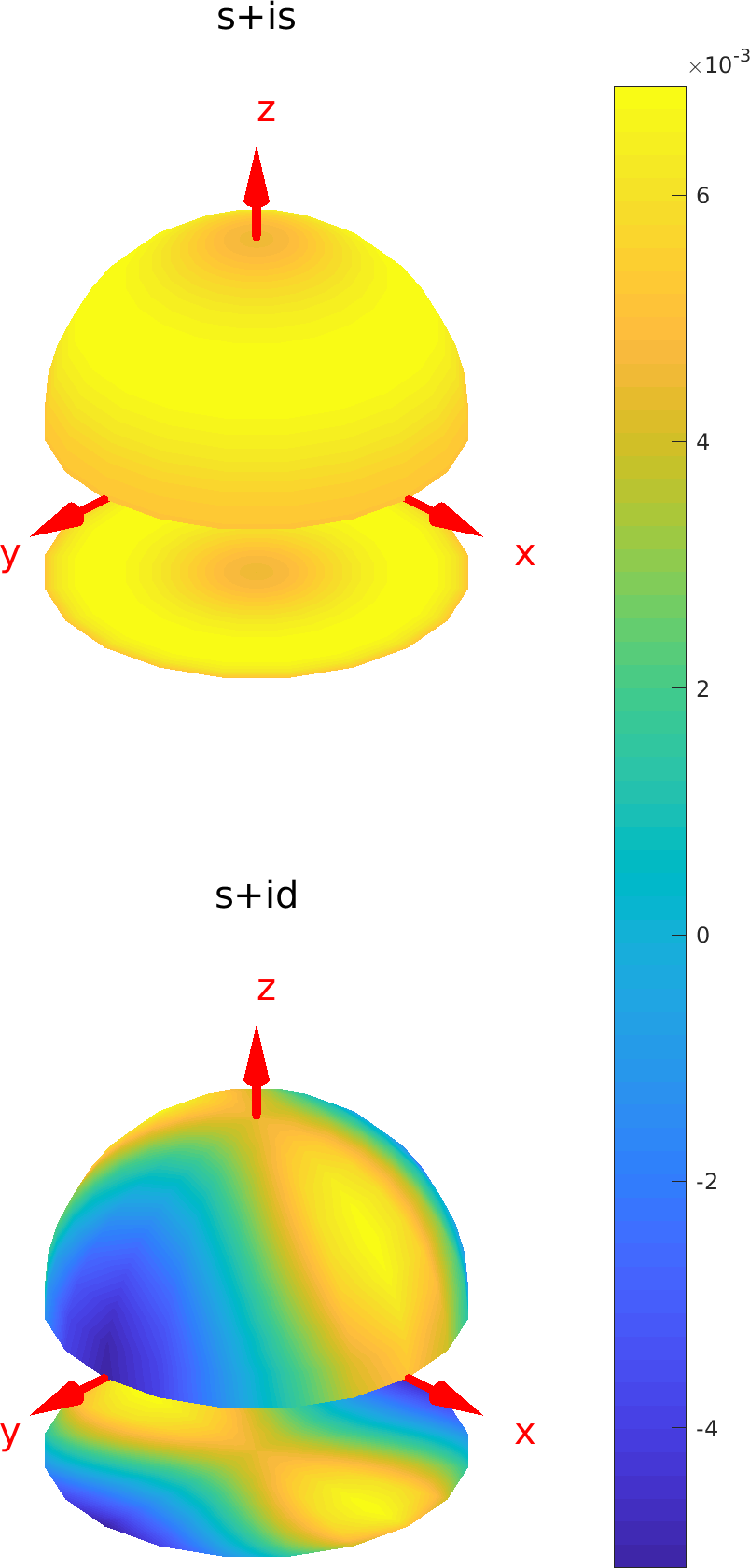}
			\end{minipage}
	\caption{Plots of the value of $Q^{12}_{11}$ and the total free energy difference between the clockwise and anticlockwise domain wall solutions for both $s+is$ and $s+id$ for all orientations. We have mapped each possible orientation (normal vector)  to a point on the unit two-sphere. The sphere has then been coloured by the value of $Q^{12}_{11}$ after performing the transformation in \eqref{trid} (left panel) and the energy of a clockwise 
	domain wall minus that of an anticlockwise domain wall (right panel). We observe similar qualitative features to the two plots, for both $s+is$ and $s+id$, in particular the contours where the functions are $0$. This supports the claim that  sign of $Q^{12}_{11}$ is a good indicator for which domain wall is lower energy.}
			\label{fig:DWs_Diff}

\end{figure*}

By considering the free energy plots we can see the predicted spatial symmetries of the full three dimensional models: $SO(2)\times C_2$ for $s+is$ and $C_2 \times C_2 \times C_2$ for $s+id$.  Note that the clockwise domain wall is the minimal energy solution for all orientations in $s+is$, whereas the minimal energy solution switches between clockwise and anticlockwise solutions depending on the orientation for the $s+id$ system.  This matches the prediction of the simple model \eqref{eqn:Fred} well: it is straightforward to see that $Q^{12}_{11}>0$ for all orientations for the $s+is$ model, and the orientations where anticlockwise domain walls are favoured in the $s+id$ model match closely the orientations where
$Q^{12}_{11}<0$, see  \figref{fig:DWs_Diff}. 

\begin{figure*}
	   \centering
\textbf{\large Maximum magnetic field} \par\medskip\vspace{0.2cm}
	\begin{minipage}{.45\textwidth}
		\includegraphics[max size={\textwidth}{\textheight}]
		{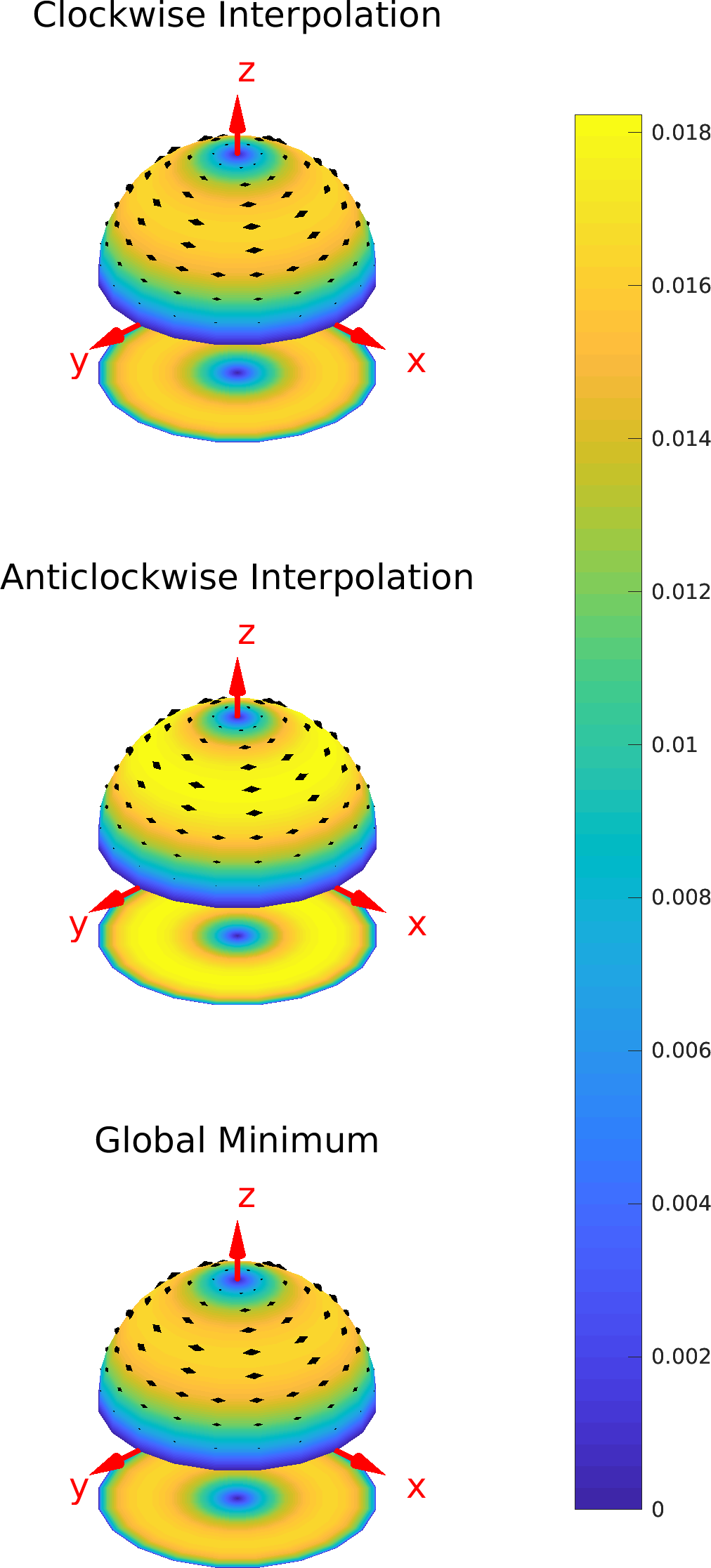}
		\subcaption{ $s+is$} \label{fig:DWs_B_s}
	\end{minipage}%
	\hfill
	\centering
	\begin{minipage}{.45\textwidth}
		\includegraphics[max size={\textwidth}{\textheight}]
		{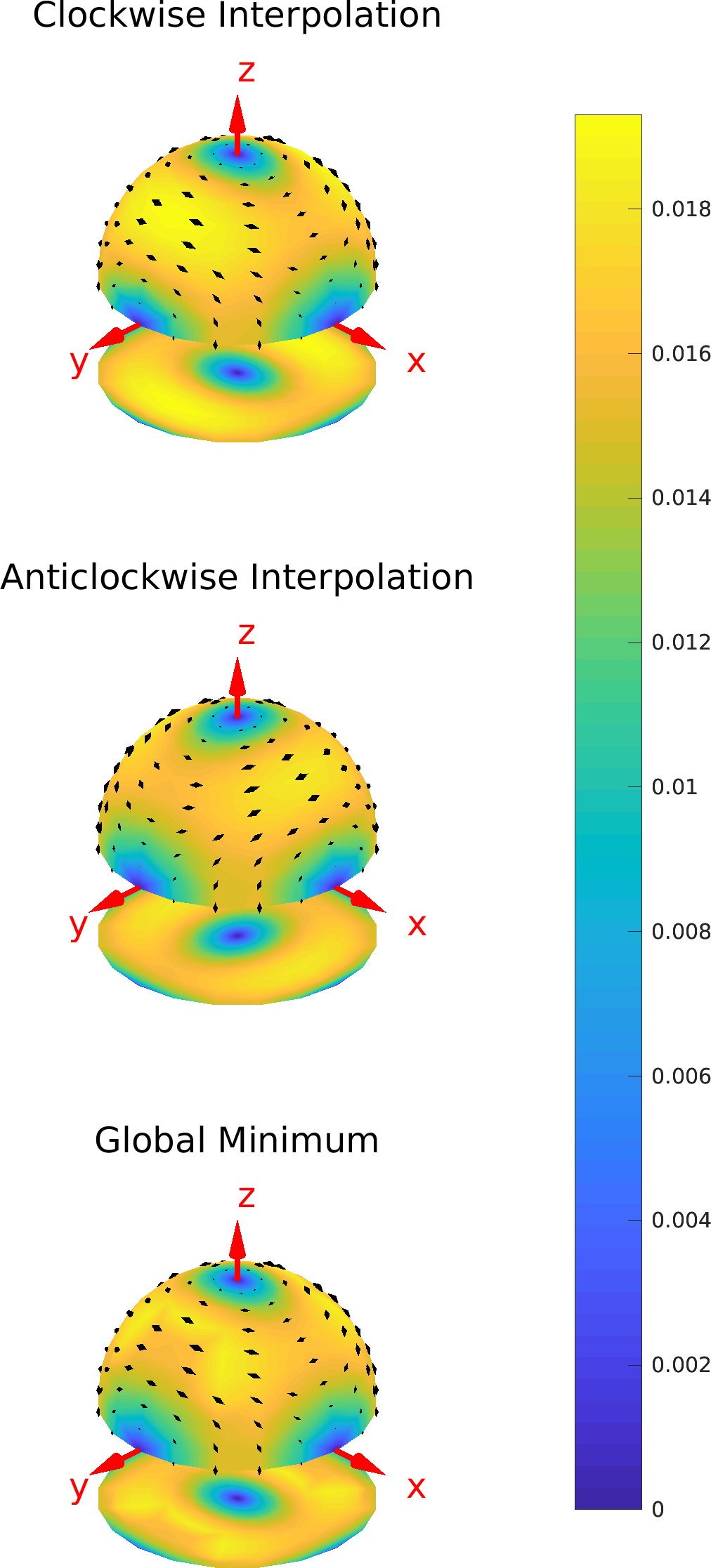}
		\subcaption{$s+id$}
		\label{fig:DWs_B_d}
	\end{minipage}
	\caption{Plot of the spontaneous field strength of each domain wall solution for all possible orientations.  We have mapped each possible orientation (normal vector) to a point on the unit two-sphere. The sphere has then been coloured by the maximum magnetic field strength of the corresponding domain wall solution.  In addition, we have plotted the direction of the local spontaneous field, where its strength is a maximum, as an arrow tangent to the 2-sphere. There are two non-degenerate domain wall solutions depending on the winding of the phase difference (clockwise and anticlockwise). This plot matches the prediction made in the linearization section and offers an experimentally verifiable signature. The spontaneous field strengths are 2 orders of magnitude lower than $H_{c_2}$ in the basal plane.  }
\end{figure*}

\begin{figure*}
	\includegraphics[max size={\textwidth}{\textheight}]
	{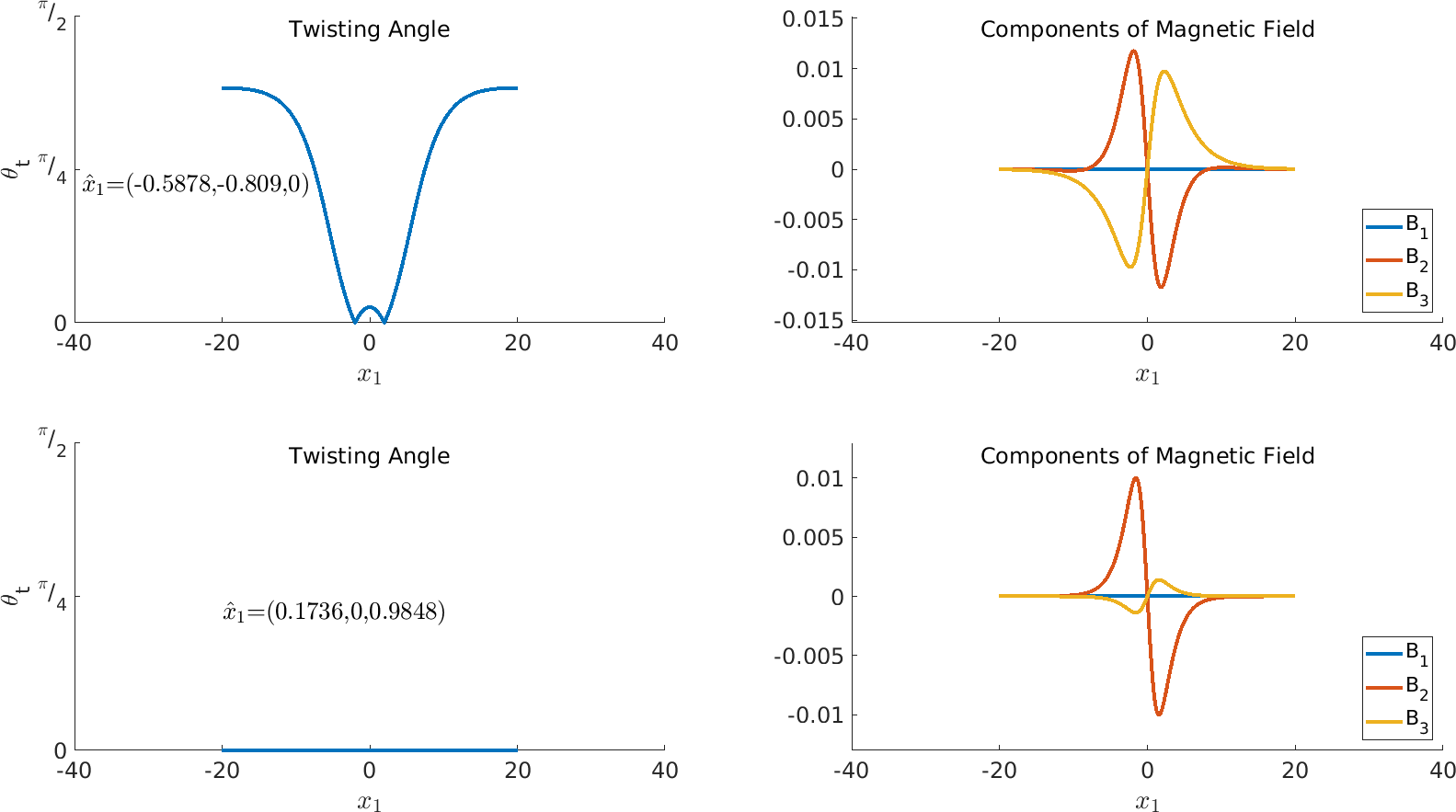}
	\caption{
	Plots of the twisting angle $\theta_t(x_1)$, for a clockwise $s+id$ domain wall. This shows how the direction of the spontaneous magnetic field changes as it decays from the point where the spontaneous field is strongest. The top row corresponds to the orientation that exhibits the most twisting $\hat{x_1} = (-0.5878,-0.8090,0)$. The bottom row corresponds to one of many orientations with no twisting $\hat{x_1} = (0.1736,0,0.9848)$, such that the spontaneous magnetic field is in the same direction at all points in space. Note that for $s+is$ no orientations exhibit spontaneous magnetic field twisting for domain walls.} \label{fig:DWs_1D_Twisting_d}
\end{figure*}

In addition, the corresponding maximum magnetic field strength is plotted for all orientations in \figref{fig:DWs_B_s} and \figref{fig:DWs_B_d}. We have also added arrows showing the direction of the maximal magnetic field (which are always tangent to the surface of the hemisphere). By this, we mean the direction of the spontaneous field at any point $x_1$ where $|B(x_1)| = B_{max}$. The spontaneous field $B(x_1)$ is an odd function about the centre of the domain wall (assumed to be $x_1 = 0$) which is consistent with the topological requirement that $\int_{x_1} B(x_1) dx_1 = 0$. This means there will be two points $\pm x_1^{max}$ where the spontaneous field corresponds in magnitude to $B_{max}$ with opposite magnetic field $B(\pm x_1^{max}) = \pm B_{max}$.  An example of this can be seen in  \figref{fig:DWs_1D_Twisting_d}, where the different components of the spontaneous magnetic field are odd functions about the centre of the domain wall $x_1 = 0$.  Hence, all the plotted arrows are double sided, representing this symmetry of the solutions.

The plots of maximum magnetic field demonstrate that there is no spontaneous field generation when the normal is aligned with any of the crystalline axes, as predicted by the linearization, which has no mixed modes for such orientations.  In addition, the spontaneous field direction on great circles (where $B_{max} \neq 0$) connecting crystalline axes (e.g.  the great circle $\hat{x}_1 = (\cos \omega, 0, \sin \omega)$ for $\omega \in [0,2\pi]$), matches the prediction from the linearized theory. In particular, the linearized theory predicts a single direction of spontaneous magnetic field orthogonal to the great circle,  as is seen in the full nonlinear numerical solutions. Note that the great circle corresponding to the basal plane for $s+is$ exhibits no spontaneous field $B_{max} = 0$ (as predicted), where as for $s+id$ there is spontaneous field. Hence, for $s+id$ this creates a vorticity in the tangent arrows $\hat{B}_{max}$ about each of the crystalline axes (where $B_{max} = 0$).  If we visualise the spontaneous maximum magnetic field as a continuous vector field on $S^2$, then we can characterize how the field circulates a given crystalline axis using a winding number $N$. Hence, if $B_{max}$ rotates clockwise once ($N = 1$) or anticlockwise once ($N = -1$) as we circle the axis. The crystalline axes at the north and south pole both have $N=1$ for both domain wall solutions.  However, the clockwise/anticlockwise domain wall solutions have $N = +/-$ about the $\hat{y}$-axis and $N = -/+$  about the $\hat{x}$-axis respectively.

Finally,  we consider how the spontaneous magnetic field locally twists direction as $x_1$ increases. We compare the spontaneous field direction with  that of $B_{max}$, defining the local twisting angle to be,
\begin{equation}
\tan \theta_t(x_1) = \frac{ |\boldsymbol{B}_{max}   \times \boldsymbol{B}(x_1) |}{ |\boldsymbol{B}_{max} \cdot \boldsymbol{B}(x_1)|}.
\end{equation}
Note that while there are two values of $x_1$ that correspond to $|B_{max}|$ with magnetic field $\pm |B_{max}|$, the chosen point has no effect on $\theta_t$.  The spontaneous field and twisting angle are plotted for two different orientations for a clockwise $s+id$ domain wall in \figref{fig:DWs_1D_Twisting_d}.  We note that the $s+is$ solutions exhibits no twisting for all orientations, which matches the linearization.  This is due to all orientations $\hat{x}_1$ for $s+is$, having at most a single direction of magnetic field for any mixed mode. This does \emph{not} mean that there is only a single mixed mode for the given orientation, but that all mixed modes share the same magnetic field direction as given in \eqref{eq:linBdir}. Hence, this predicts that all spontaneous magnetic field will be in the same direction and exhibit no twisting.  The nonlinear solutions for all orientations match this prediction, exhibiting no twisting and with all spontaneous fields matching the predicted linear direction.

$s+id$ models, in contrast, exhibit significant magnetic field twisting as can be seen in \figref{fig:DWs_1D_Twisting_d}. This is a result of a different mixed mode dominating in the nonlinear region of the domain wall (where $|B|$ is large) and the linear region (when $|B|$ is small), causing the spontaneous magnetic field to twist direction as it decays from its maximum value ($\theta_t = 0$). Note that as a result of the topological requirement $\int B dx_1 = 0$, the magnetic field is an odd function about the centre ($x_1 = 0$) as can be seen in the plots of the different magnetic field components. 

\begin{figure*}
	\centering 
	\textbf{\large Maximum Twisting Angle ($\theta^{max}_t$)  }  \par\medskip\vspace{0.2cm}
	\includegraphics[max size={\textwidth}{\textheight}]
{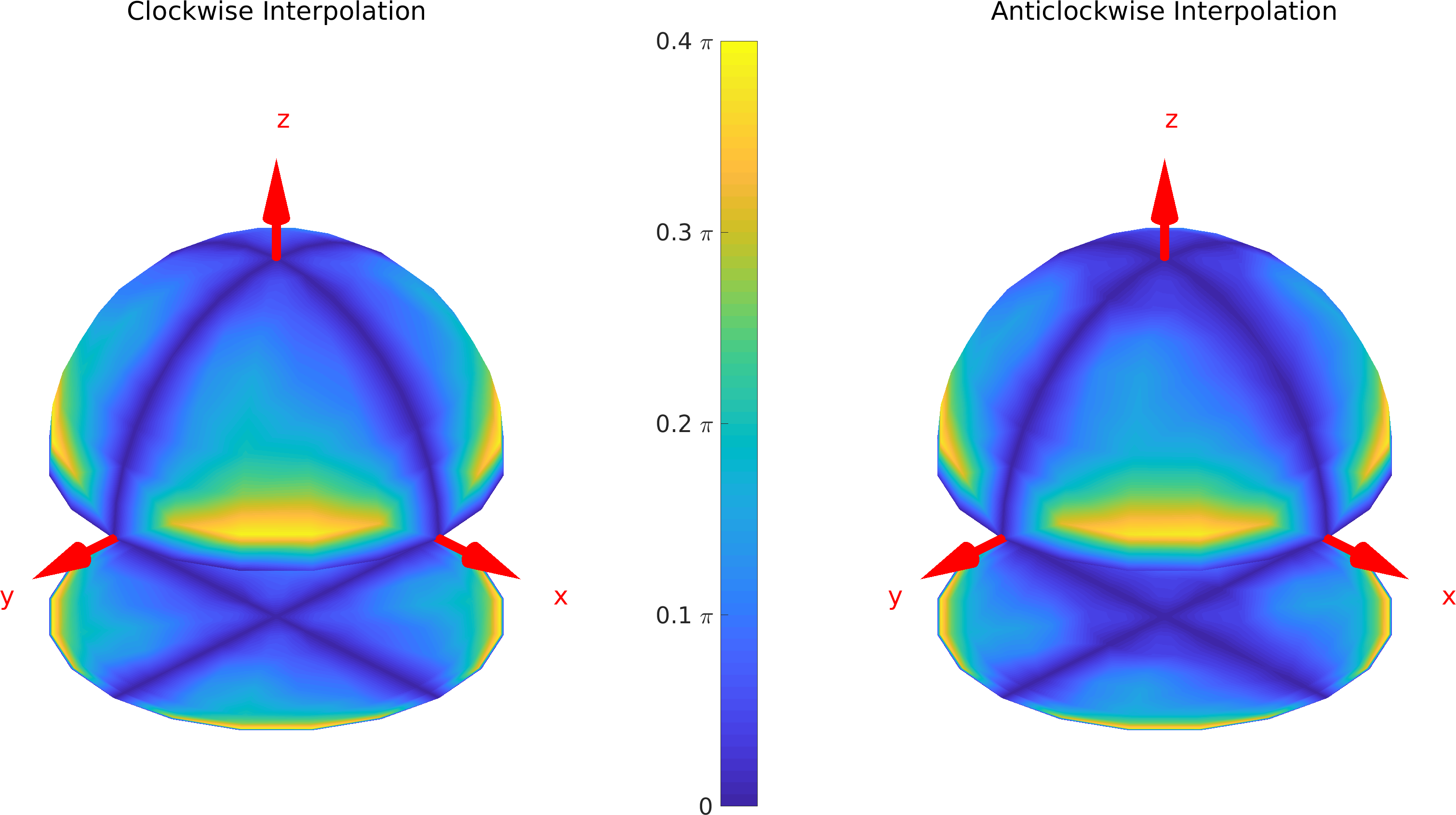}
\caption{Plots of the maximum twisting angle  of the magnetic field of domain wall in an $s+id$ superconductor: clockwise phase difference winding (left) and anticlockwise phase difference winding (right). We have mapped each possible orientation (normal vector) to a point on the unit two-sphere. The sphere has then been coloured by the value of $\theta_t^{max}$ for that orientation. The twisting ( $\theta_t^{max}$ ) is the same for the two domain wall types.} \label{fig:DWs_T_d}
\end{figure*}

To demonstrate how the amount of twisting for $s+id$ models changes with orientation, we have plotted $\theta_t^{max} := \max\{\theta_t(x_1):x_1\in\R\}$ in figure \figref{fig:DWs_T_d}. This shows that on the great circles that connect crystalline axes there is no twisting, which matches the linearization. Like with $s+is$ models, on these great circles the linearization predicts a single direction for the magnetic field for all mixed modes and hence no twisting. However, away from these great circles the twisting becomes significant for both the clockwise and anticlockwise domain walls. This offers an experimental signature that can differentiate between $s+is$ and $s+id$ systems. 

In summary,  domain walls produce spontaneous magnetic fields due to mode mixing. This is due to the anisotropy of the model, causing the modes to have both matter and magnetic components. While the linear modes are only strictly justified far from the excitation of the domain wall, we have demonstrated that they are remarkably accurate at predicting the spontaneous magnetic fields even when the model is nonlinear dominated. This suggests that spontaneous fields for anisotropic models can be predicted accurately using the linearized model alone. This is quite a remarkable feature of a traditionally highly nonlinear model.  In addition, we have demonstrated that both $s+is$ and $s+id$ models exhibit two different domain wall solutions (coined clockwise and anticlockwise solutions). Finally $s+id$ models exhibit significant magnetic field twisting as the spontaneous fields decay $x_1 \rightarrow \infty$.

\section{Upper critical field}
For mathematical convenience, we have worked throughout with dimensionless quantities. To get a rough idea of the size of the spontaneous magnetic fields predicted in real systems, it is useful to compare $B_{max}$ with the upper critical field
$H_{c_2}$ for the systems studied. This may be computed numerically using the standard strategy (reducing the GL equations linearized about the normal state to a coupled harmonic oscillator problem). Note that $H_{c_2}$ is anisotropic, that is, it depends on the direction of the applied field.

 We find that the value of $H_{c_2}$ for $s+is$ varies between $1.6468$ ($H$ parallel to the basal plane) and $4.2656$ ($H$ in $z$-direction) and has $SO(2)$ symmetry about the $z$-axis, as it must.
 
For $s+id$, we find that $H_{c_2}$ matches the four fold symmetry about the $z$-axis of the free energy and is maximal in the $z$-direction with $H_{c_2}=2.6596$ and minimal in the basal plane, going as low as $H_{c_2} = 1.0245$.

The key takeaway from this calculation is that the spontaneous fields from the previous section are approximately two orders of magnitude weaker then $H_{c_2}$ in the basal plane. This is strong enough to be detected using multiple experimental techniques. 

\section{Conclusion}

In conclusion,  we have demonstrated that the familiar London model is, in general, not accurate in describing the magnetic field behaviour of $s+is$ and $s+id$ systems. This is a consequence of the normal modes not separating into purely magnetic and matter modes, but being mixed. 
This means even a small pertubation of the superconducting gap induces magnetic field and vice versa.
This mixing of both magnetic and matter components was shown to be a generic feature of anisotropic models with mixed gradient terms.

The key observable consequence of mixed modes is their contribution to spontaneous magnetic fields.
The orientation dependence of these spontaneous fields gives an experimentally verifiable signature for the pairing symmetries of the system. This also explains the previous results in \cite{benfenati2020magnetic}, and allows spontaneous field directions to be predicted using linear algebra.

In addition, we have extended the previous approach to linearizing anisotropic models  \cite{speight2019chiral}. We demonstrated that the familiar symmetry reductions used to study 1-dimensional excitations are not always valid in these systems. In particular, due to the anisotropy one must include all components of the vector gauge field $A_i$, else the symmetry reduction will not, in general, be a solution of the full 3-dimensional equations of motion. It is this approach that allows the magnetic field to twist direction as it decays.

We also demonstrate that, in general, the behaviour of the magnetic field cannot be characterized by a single length scale: the London magnetic field penetration length.

For $s+is$ models the decaying field in the Meissner state, exhibits magnetic field twisting, due to modes that spontaneously generate magnetic field, orthogonal to the applied field direction. Instead, various components of the magnetic field decay with different length scales. Hence, as the fields decay, the magnetic field twists towards the mode with the longest length scale. However, for $s+is$ excitations that exhibit purely spontaneous magnetic fields, there is no field twisting, due to all mixed modes having equivalent magnetic components.

$s+id$ models, in contrast, exhibit twisting due to both disparate length scales and purely spontaneous fields. This is due to the $s+id$ mixed modes having multiple magnetic field directions, such that the spontaneous fields twist as they decay. This has been shown to result in significant magnetic field twisting for domain walls, and it is likewise expected to occur for defects.

The spontaneous magnetic fields for domain walls in both $s+is$ and $s+id$ were studied in detail. These spontaneous fields offer one of the best experimental signatures to differentiate between various pairing symmetries. This can be achieved using scanning probes of magnetic fields of domain walls, pinned in various orientations relative to crystal axes.

\section{Acknowledgements}
We thank Andrea Benfenati and Mats Barkman for useful discussions. The work of MS, TW and AW is supported by the UK Engineering and Physical Sciences Research Council through grant EP\/ P024688\/ 1 (MS and TW)
and a research studentship (AW). TW is also supported by an academic development fellowship, awarded by the University of Leeds. EB is supported by the Swedish Research Council Grants No. 2016-06122, 2018-03659 and G\"{o}ran Gustafsson  Foundation  for  Research  in  Natural  Sciences  and  Medicine
and Olle Engkvists Stiftelse. The numerical work of this paper was performed using the code library Soliton Solver, developed by one of the authors, and was undertaken on ARC4, part of the High Performance Computing facilities at the University of Leeds.

\begin{appendices}

\section{Parameters Used} \label{app:parameters} 
All simulations make use of the following potential,
\begin{align}
F_P = -&\frac{1}{2}|\psi_1|^2 -\frac{1}{2}|\psi_2|^2 + 2|\psi_1|^4 + 3|\psi_2|^4 + \frac{3}{2} |\psi_1|^2|\psi_2|^2 \nonumber \\ + & \frac{1}{8}|\psi_1|^2|\psi_2|^2 \cos{2\theta_{12}},
\end{align}
where we have set $\alpha_\alpha = -1/2$, $\beta_1 = 4$, $\beta_2 = 6$, $\gamma = 3/2$ and $\eta = 1$. In addition, the anisotropy matrices $Q^{\alpha\beta}$ are set as given in table \ref{Tab:Q2}, where we have set $a_1 = 4$, $a_2 = 1/2$, $a_3 = 1$, $b_1 = 0.3$, $b_2 = 2$ and $b_3 = 0.2$ for both $s+is$ and $s+id$ models.
\begin{table}[H]
	\begin{center}
		\begin{tabular}{c|c} 
			\hline
			s+is & s+id\\
			\hline
			$Q^{11} = \left(\begin{array}{ccc} 4 & 0 & 0 \\ 0 &4 & 0 \\ 0 & 0 &0.3 \end{array}\right)$ &  $Q^{11} = \left(\begin{array}{ccc} 4 & 0 & 0 \\ 0 &4 & 0 \\ 0 & 0 &0.3 \end{array}\right)$\\
			$Q^{22} = \left(\begin{array}{ccc} 0.5 & 0 & 0 \\ 0 &0.5 & 0 \\ 0 & 0 &2 \end{array}\right)$ &  $Q^{22} = \left(\begin{array}{ccc} 0.5 & 0 & 0 \\ 0 &0.5 & 0 \\ 0 & 0 &2 \end{array}\right)$\\
			$Q^{12} = \left(\begin{array}{ccc} 1 & 0 & 0 \\ 0 & 1 & 0 \\ 0 & 0 &0.2 \end{array}\right)$ &  $Q^{12} = \left(\begin{array}{ccc} 1 & 0 & 0 \\ 0 & -1 & 0 \\ 0 & 0 &0.2 \end{array}\right)$\\
			\hline
		\end{tabular}
	\end{center}
	\caption{Form of the anisotropy matrices for the simulated $s+is$ and $s+id$ systems.}
	\label{Tab:Q2}
\end{table}

\section{Natural Boundary Conditions}
To find numerical solutions of the Meissner state in the region $\Omega$ we must minimize the Gibbs free energy in \eqref{Eq:Gibbs} among all fields $\phi_a$, $a\in[1,6]$ defined on $\Omega$.  This leads to the following variation for $G$,
\begin{align}
\delta G = & \int_\Omega \left\{ \frac{\partial \mathcal{G}}{\partial \phi_a } - \partial_i\left(\frac{\partial \mathcal{G}}{\partial(\partial_i \phi_a)}\right)\right)\delta\phi_a  \\
+ & \int_{\partial\Omega} \left(\frac{\partial \mathcal{F}_{surf}}{\partial \phi_a} - n_i \frac{\partial \mathcal{G}}{\partial(\partial_i \phi_a)}\right) \delta \phi_a,
\end{align}
where we have used the divergence theorem, and recalled that $\nvec$ is an inward pointing normal to $\partial \Omega$. Demanding that $\delta G = 0$ for all variations requires both of these integrals to vanish identically and hence $\phi_a$ satisfies the usual Euler-Lagrange equations in $\Omega$ together with the boundary conditions,
\begin{align}
n_i Q^{1\beta}_{ij} D_j \psi_\beta = 0,\\
n_i Q^{2\beta}_{ij} D_j \psi_\beta = 0,\\
\partial_i A_i = 0,\\
B = H.
\end{align}
This can be simplified by first performing a change of basis from the crystaline basis $(\hat{x},\hat{y},\hat{z})$ to the excitation basis $(\hat{x}_1, \hat{x}_2, \hat{x}_3)$ by performing the transformation in \eqref{trid} on the anisotropy matrices. This leads to the following simpler boundary conditions in the new basis,
\begin{align}
Q^{1\beta}_{11} D_1 \psi_\beta = 0,\\
Q^{2\beta}_{11} D_1 \psi_\beta = 0,\\
A_1' = 0,\\
B = H.
\end{align}
We impose these boundary conditions at $x_1 = 0$, then at $x_1 = L$, where $L$ is large, we demand that $b' = a' = 0$, $\psi_1 = u_1$ and $\psi_2 = iu_2$, such that the fields are in their ground state. 

Note, as we are interested in bulk behaviour, we have neglected the presence of surface terms \cite{samoilenka2020microscopic}, that lead to additional magnetic effects \cite{benfenati2021spontaneous}.

\end{appendices}

\bibliography{reference}{}

\end{document}